\documentclass{aastex}   
\usepackage{emulateapj5}

\def\XMM{{\it XMM-Newton}}
\def\ltsima{$\; \buildrel < \over \sim \;$}
\def\simlt{\lower.5ex\hbox{\ltsima}}
\def\gtsima{$\; \buildrel > \over \sim \;$}
\def\simgt{\lower.5ex\hbox{\gtsima}}
\newcommand{\eqn}[1]{(\ref{#1})}
\def\gr{\kern 2pt\hbox{}^\circ{\kern -0.5pt K}} 

\shorttitle{Cooling and heating in M87}
\shortauthors{Ghizzardi \it{ et al.}}

\begin{document}

\title{Radiative cooling, heating and thermal conduction in M87}

\author{Simona Ghizzardi, Silvano Molendi, Fabio Pizzolato\altaffilmark{1}}
\affil{Istituto di Astrofisica Spaziale e Fisica Cosmica, 
IASF -- CNR, Sez. di Milano,\\
via Bassini 15, I-20133, Milano, Italy}
\email{simona@mi.iasf.cnr.it, silvano@mi.iasf.cnr.it, fabio@mi.iasf.cnr.it} 

\and 
\author{Sabrina De Grandi} 

\affil{INAF - Osservatorio Astronomico
  di Brera, via Bianchi 46, I-23807 Merate (LC), Italy}

\email{degrandi@merate.mi.astro.it}

\altaffiltext{1}{Dipartimento di Scienze, Universit\`a dell'Insubria,
  via Valleggio 11, 22100, Como, Italy}

\begin{abstract}
  The crisis of the standard cooling flow model brought about by {\it
    Chandra} and {\it XMM-Newton} observations of galaxy clusters, has
  led to the development of several models which explore different
  heating processes in order to assess if they can quench the cooling
  flow.  Among the most appealing mechanisms are thermal conduction
  and heating through buoyant gas deposited in the ICM by AGNs. We
  combine Virgo/M87 observations of three satellites ({\it Chandra},
  {\it XMM-Newton} and {\it Beppo-SAX}) to inspect the dynamics of the
  ICM in the center of the cluster.  Using the spectral deprojection
  technique, we derive the physical quantities describing the ICM and
  determine the extra-heating needed to balance the cooling flow
  assuming that thermal conduction operates at a fixed fraction of the
  Spitzer value. We assume that the extra-heating is due to buoyant
  gas and we fit the data using the model developed by Ruszkowski and
  Begelman (2002). We derive a scale radius for the model of $\sim 5$
  kpc, which is comparable with the M87 AGN jet extension, and a
  required luminosity of the AGN of a $few \times 10^{42} {\rm erg \,
    s^{-1}}$, which is comparable to the observed AGN luminosity.  We
  discuss a scenario where the buoyant bubbles are filled of
  relativistic particles and magnetic field responsible for the radio
  emission in M87. The AGN is supposed to be intermittent and to
  inject populations of buoyant bubbles through a succession of
  outbursts.  We also study the X--ray cool component detected in the
  radio lobes and suggest that it is structured in blobs which are
  tied to the radio buoyant bubbles.
    
\end{abstract}

\keywords{conduction --- cooling flows --- galaxies: active ---
  X-rays: galaxies: clusters --- galaxies: clusters: individual
  (Virgo)}

\section{Introduction}

The hot diffuse X-ray emitting gas (intracluster medium, ICM for
short) provides a powerful tool to inspect the internal dynamics of
galaxy clusters.  For the typical density and temperature of the
intracluster gas, the main emission mechanism is the bremsstrahlung
and, for a large amount of clusters, the radiative cooling time in the
central regions is significantly shorter than the Hubble time.  As a
consequence, if no additional heating mechanism is present, the gas
cools and is expected to flow inwards, forming a {\it cooling flow}.
The standard model of cooling flows \citep[see][for a review]{fa94}
predicted the gas to be a multiphase medium in which there is a broad
range of temperatures and densities present at all radii.  Mass
deposition rates were estimated to be as large as hundreds of solar
masses per year \citep{Allen01}.  This model was strengthened by the
general thought that in presence of magnetic fields the thermal
conduction must be highly suppressed \citep{BC81,fa94,CC98, Mal01},
which is a necessary condition for the multiphase cooling to operate.
In fact, no heating exchange between the different phases must occur
in order that they may coexist.  There is some observational evidence
that modest magnetic fields are present throughout the intracluster
medium.  The current measurements of intracluster magnetic fields are
based on Faraday rotation measure (RM) in radio sources seen through
clusters \citep[e.g.][]{kim91,CKB,Fer99,Tay01}; direct evidence also
comes from measurements of extended regions of radio synchrotron
emission in clusters \citep[see e.g.][]{GF00,fusco,OMV,F99}. Both the
excess RM values and the radio halo data suggest modest magnetic
fields, at a few microgauss levels, throughout the cluster.

Recent \XMM\, and {\it Chandra} observations have shown that in the
central regions, the temperature drops to about one third of its
overall mean value and there is no evidence of temperatures smaller
than $\sim 1-2$ keV \citep{Pet01,Kaa01,Tam01,Allen01}, suggesting that
the gas does not cool below these cutoff temperatures.  Moreover, the
new estimated mass deposition rates are significantly smaller than
those evaluated by using previous X-ray satellites data
\citep{McN01,Pet01}.  Lastly, the new data show that clusters spectra
are better represented by a single (or double) temperature model
rather than the standard multiphase (multi-temperature) cooling flow
model \citep{MP01,Bohr01,Fab01,matsu}.

These new results clearly show that the standard cooling flow model is
not a satisfying description of the internal dynamics of the ICM.
Some source of heat which stops the cooling flow and balances
radiative losses must be sought.  The nature of this source and the
origin of the heat mechanism is still unclear.

One possible candidate is thermal conduction.  Recent works by
\citet{NM01} and \citet{gruz02} show that in the presence of turbulent
magnetic fields, the conductivity can be as large as a fraction of the
Spitzer value and thus can play a significant role in balancing
cooling flows. As a consequence, thermal conduction has been recently
re-introduced as a possible heat source to balance the energy losses
\citep[see e.g.][]{Voigt,VF04,Fab02, Mal01,ZN02}.  However, as we will
discuss more in detail in \S \ref{sub:cond}, thermal conduction fails
in supplying the needed heat in the central regions \citep[see
also][]{Voigt,VF04, ZN02}.

Heating from a central active galactic nucleus is another possibility.
The idea is supported by the fact that most of ``cooling clusters''
host a central active galactic nucleus with strong radio activity
\citep{Burns,BM02}. Several models in the literature explore AGN
heating processes to assess if they can balance the radiative losses.
One of the most appealing mechanisms involves buoyant gas bubbles,
inflated by the AGN, that subsequently rise through the cluster ICM
heating it up \citep{ch02,Bohr02,BKa,BKb,Brug02}.

However, all the models predicting that radiative cooling is balanced
only by energy input from the central AGN fail
\citep{McN02,ZN02,BM02}.  In particular, \citet{BM02} analyzed several
heating mechanisms induced by the central AGN and concluded that no
simple mechanism is able to quench the cooling flow.  Moreover, the
required mechanism needs a finely tuned heating source. Indeed, the
heat source must provide sufficient energy to stop the cooling flow,
but not enough to trigger 
strong convection or the metallicity gradients observed in all
cooling flow clusters \citep{DM01} would be destroyed. As a
consequence, an AGN can be an efficient mechanism in the very center
of the cluster but it is unlikely to be strong enough to provide
energy to the outer parts of the ``cooling region''. So, it may be
viewed as complementary to thermal conduction which fails in quenching
the cooling flow in the innermost regions.

Recently, \citet{RB02} and \citet{ZN02} concluded that both thermal
conduction and heating from a central AGN can play an important role
in balancing the cooling.  In particular, \citeauthor{RB02} (RB02
hereafter) developed a model where both thermal conduction and heating
from a central AGN co-operate in balancing the radiative losses.  One
of the main advantages of this model is that it reaches a stable final
equilibrium state and it is able to reproduce the main observed
quantities, such as the temperature profile (with a minimum
temperature $T \sim 1$ keV).

In this paper, we use M87/Virgo observations of three satellites
(namely \XMM, {\it Chandra} and {\it Beppo-SAX}) to test various
heating models on this cluster. To this end, we apply to the M87 data
the deprojection technique to recover some physical quantities of the
ICM such as the gravitational mass, the entropy and the heating
required to balance the cooling flow.

The paper is organized as follows: in \S \ref{sec:sp2d} we report
details about the analysis of the three ({\it Chandra}, \XMM\, and
{\it Beppo-SAX}) M87 datasets; in \S \ref{sec:depro}, we revise
briefly the spectral deprojection technique that we adopt for our
analysis; in \S \ref{sec:m87}, we deproject the M87 data, we test that
the spherical symmetry hypothesis holds and we derive the
gravitational mass for M87; in \S \ref{sec:heat}, we determine the
amount of extra-heating needed to balance the cooling flow when
thermal conductivity is assumed to operate at a fraction of the
Spitzer value.  Lastly, in \S \ref{sec:concl} we summarize our
results.

\section{Data analysis}
\label{sec:sp2d}

Thanks to its proximity, the Virgo cluster and its giant elliptical
central galaxy M87, represent an incomparable target to inspect the
internal properties of the ICM.  Aiming to a precise characterization
of the ICM, we use observations of the three satellites {\it Chandra},
\XMM\, and {\it Beppo-SAX}.  These satellites views are complementary:
the sharp PSF ($\sim 0.5''$) of {\it Chandra} can provide a precise
analysis of the innermost (say $\simlt 5-10$ kpc) regions of M87; the
\XMM\, large collecting area and its wide field of view allow a good
inspection of the intermediate regions (up to $ \sim 80$ kpc); in the
outermost regions of the cluster, where the angular resolution is less
critical and \XMM\, data are highly contaminated by the background,
{\it Beppo-SAX} is a better choice and data can be collected out to a
radius of $\sim 120$ kpc.

\subsection{\XMM\, data preparation.}

M87 has been observed during the PV phase of \XMM.  The details of
this observation have been widely discussed in several publications
\citep[see e.g.][]{Bohr01,belsole,MP01,MG01,GM02,matsu}.  We make use
of the results of the spectral analysis described and discussed in
\citet[M02 hereafter]{ms02}.  The cluster is divided in 139 regions
for 12 concentric annuli, centered on the emission peak.  The regions
are the same as those presented in M02, apart from the annuli in the
$1-4$ arcmin range which have been taken 0.5 arcmin wide instead of 1
arcmin wide.  Unlike \citet{matsu}, we decided to use annuli at least
$30''$ wide in order to avoid possible PSF contaminations \citep[see
also][]{Mark02}. For a detailed description of the MOS PSF see
\citet{PSF1}.

We refer the reader to M02 for all the details of the spectral
analysis procedure and remind the reader that the accumulated spectra
in all the regions are fitted with two different models: (i) a single
temperature (1T) model ({\it vmekal} in {\sc{xspec}}) and (ii) a two
temperature (2T) model ({\it vmekal} + {\it vmekal} in {\sc{xspec}}).

The 1T fits of the accumulated spectra provide for each region the
emission-weighted temperature $T$ of the gas and the emission integral
$EI=\int{n_e n_pdV}$ where $n_e$ and $n_p$ are the electron and proton
density.  The 2T fits of the spectra provide for each region the
temperatures and the emission integrals of the two different
components of the gas. These quantities will be used to derive the
density and the temperature profile of the cluster.

\subsection{{\it  Chandra} data preparation.}

We have analyzed the {\it Chandra} ACIS-S3 observation \citep[obs.
id.  352; see also][]{Young} centered on M87
($\alpha=12$:$30$:$49.40$; $\delta=+12$:$23$:$27.82$) using CIAO 2.1.1
and CALDB 2.15.  We have followed the procedures described in the
Science Threads available at the Chandra X-ray Center on-line pages.
The light curve was filtered for high background events obtaining an
effective exposure time of 35.2 ks.  From our analysis we excluded the
AGN and the associated jet cutting off a narrow rectangular region
centered in $\alpha=12$:$30$:$48.80$, $\delta=+12$:$23$:$31.37$
(J2000) with lengths $\Delta \alpha=25^{\prime\prime}$, $\Delta
\delta=5^{\prime\prime}$ and rotated by $22^o$ from E to N.  We have
extracted spectra from regions of concentric annuli centered on the
emission peak ($0^{\prime\prime} - 10^{\prime\prime}$,
$10^{\prime\prime} - 30^{\prime\prime}$, $30^{\prime\prime} -
45^{\prime\prime}$, $45^{\prime\prime} - 60^{\prime\prime}$,
$60^{\prime\prime} - 80^{\prime\prime}$, $80^{\prime\prime} -
100^{\prime\prime}$ and $100^{\prime\prime} - 120^{\prime\prime}$);
the $10^{\prime\prime} - 30^{\prime\prime}$ annulus was divided into
four 90$^o$ regions starting from a position angle of 45$^o$, all the
other annuli have been divided into 8 regions 45$^o$ wide starting
from position angle 0$^o$.  The background used in the spectral fits
was extracted from blank-sky observations using the {\it
  acis-bkgrnd-lookup} script.  We have fitted each spectrum with the
same models (1T and 2T) used for the analysis of the \XMM\, spectra
and using the effective areas and response matrix derived with the
routines {\it mkwarf} and {\it mkwrmf} for extended sources given in
CIAO.  The energy range is the same as the one used for the \XMM\,
spectral analysis.

\subsection{{\it Beppo-SAX} data preparation.}

We have analyzed the pointed {\it Beppo-SAX} observation of the Virgo
cluster (obs. id.  60010001), adding the MECS2 and MECS3 data and
obtaining an effective exposure time of 25.1 ks.  The analysis of data
follows the procedure described in details in \citet{DM01}. We have
extracted spectra for 7 concentric annuli centered on the emission
peak ($\alpha=187.6992$ deg, $\delta=12.3878$ deg J2000), each annulus
is 2$^\prime$ wide up to 10$^\prime$ from the peak and 4$^\prime$ wide
from 10$^\prime$ to the maximum 20$^\prime$ radius.  We have fitted
each spectrum with a {\it mekal} model absorbed for the Galactic
$N_H$ using the appropriate effective area computed for extended
source as described in \citet{DM01} and the background spectrum
extracted from blank-sky files for the same annular regions.  The
energy range considered in the spectral analysis is $2.-10.$ keV in
all cases out of the $8^\prime-12^\prime$ annulus for which we have
considered the $3.5-10$ keV range to avoid spectral distortions from
the supporting structure of the instrument entrance windows.  Note
that, as we will discuss in the next paragraph, for {\it Beppo-SAX}
data, only the 1T model has been used to fit the accumulated spectra.

\subsection{The joint data set.}
\label{sub:joint}

For each region observed with \XMM\, or {\it Chandra}, an F-test is
used to establish whether the 2T model provides a better description
of the data with respect to the 1T model.  We used two different
criteria for \XMM\, and {\it Chandra} data to select the regions
represented by a 2T model.  As far as \XMM\, data are concerned, for
those regions whose F-test provides a probability $ \ge 95\%$, the 2T
description is retained, whereas for those regions whose F-test
provides a probability $ \le 90 \%$, the 1T description is adopted.
The regions having an F-test probability within the [0.90-0.95] range
have been rejected and excluded from the analysis.  A somewhat more
stringent criterion has been adopted for the {\it Chandra} data,
regions with F-test probability within the [0.75-0.98] range have been
rejected, because of the lower statistical quality of the latter
dataset.

It is worth noting that M02 shows (using \XMM\, data) that the regions
which are better represented by a 2T model match the radio ``arms''
which are visible in the M87 map at 90 cm \citep{owen}.  At large
radii, where we have no evidence of a second temperature component
from the \XMM\, data, we will use the results of 1T fits to the {\it
  Beppo-SAX} data, which, because of its limited spectral coverage, is
insensitive to the cooler component.

\begin{figure}
\epsscale{0.8}
\plotone{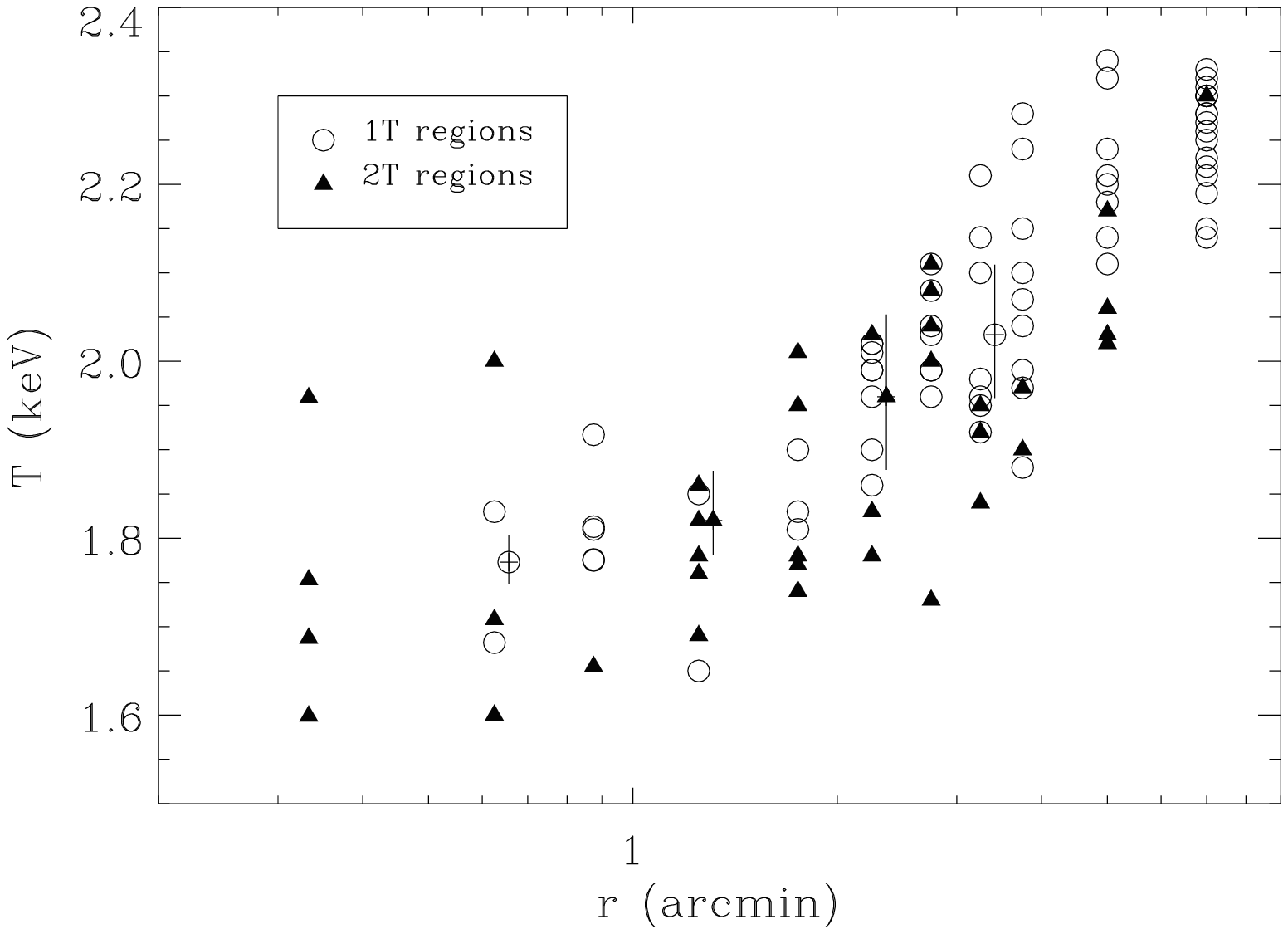}     
\caption
{ Emission weighted temperatures for the single phase gas of the 1T
  regions (open circles) and for the hot component (triangles)
  of the 2T regions.  Error bars are plotted only for few
  representative points. These points are plotted at slightly larger
  radii for a clearer view of the error bars amplitude. }
\label{fig:cfrt_s}
\end{figure}

While the cool component is related to the radio ``arms'', the hot
component in the regions described by the 2T model, is very similar to
the 1T gas of the other regions which do not feature strong radio
emission and are located at similar radial distances from the cluster
core. In Fig.  \ref{fig:cfrt_s} we plot the emission weighted
temperatures for 1T and 2T regions. The open circles represent the
temperatures of the 1T regions, while the triangles plot the
temperature of the hot component in those regions which are fitted by
a 2T model.  We plot error bars for a few representative points.  All
the other error bars are not reported for a clear reading of the plot.
The Fig.  \ref{fig:cfrt_s} shows that for those annuli where we have
1T and 2T regions, the temperature of the hot component of the 2T
regions is not separate from the temperature of the single phase gas
of the 1T regions. So, the hot component and the single phase gas are
distributed in a regular and symmetric fashion. As far as only the hot
component is considered for the 2T regions, the cluster appears to be
approximatively spherically symmetric, which is an important condition
for the application of the deprojection technique.
  
As already outlined, for each region we consider the emission integral
$EI$ and the emission-weighted temperature $T$.  For those regions
fitted by the 2T model, we retain the $EI$ and $T$ related to the hot
component.  The radially averaged profile for each physical quantity
is determined starting from the region-by-region description.  We
assign to each annulus, a mean $T$ by averaging on all the $T$ of the
regions belonging to the annulus. The $EI$ is the sum of the $EI$ of
the regions along the ring.

Some attention must be paid in this averaging (or summation for $EI$)
procedure as some portions of the observed regions can be masked. In
fact, there are some pixels of the Field of View of the different
instruments which must be rejected for different reasons: (i) they are
CCD hot or dead pixels; (ii) they correspond to the gaps between
nearby CCDs; (iii) they correspond to regions which are never observed
by the instruments (for MECS) or which are partially outside the
Field of View; (iv) they correspond to some point source contaminating
the X-ray cluster emission; (v) they correspond to the excluded AGN
rectangular regions ({\it Chandra}).  As a consequence, even if the
geometrical area (or equivalently the emitting volume) of the regions
belonging to the same ring is the same, the effective emitting area
(or volume) is a fraction of the geometrical area depending on the
number of rejected pixels.  The measured $EI$ in each region accounts
for photons coming from the effective emitting area (or volume);
hence, each $EI$ must be renormalized.  The factor of normalization is
given by the ratio $A_{geom}/(A_{geom}-A_{mask})$ where $A_{geom}$ is
the geometric area of the region and $A_{mask}$ is the total area of
the rejected pixels in the region.  The same normalization factors are
used as weights for the determination of the averaged $T$.

\begin{figure}
\epsscale{0.75}
\plotone{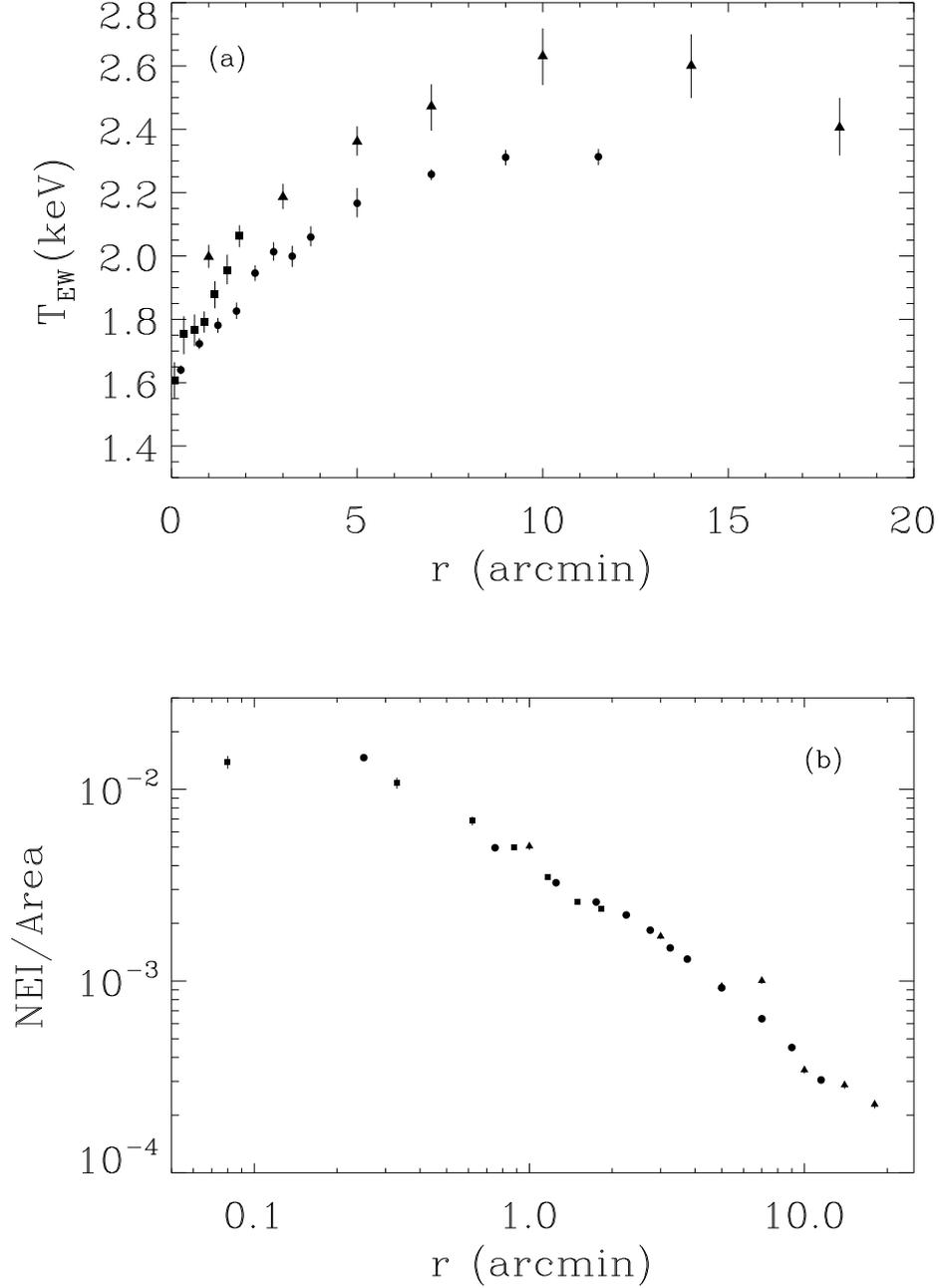}     
\caption
{(a) Emission-weighted temperature profile for M87 obtained from the
  spectral analysis described in \S \protect\ref{sec:sp2d}; (b)
  Normalized Emission Integral (NEI) profile per unit area; $NEI$ is given
  in {\sc{xspec}} units, i.e.  $NEI = {{10^{-14}} \over {4\pi
      d^2_{ang}(1+z)^2}}EI$, where $d_{ang}$ is the angular distance
  of M87 in cm, $z$ the redshift and $EI$ in cm$^{-3}$. The $Area$ is
  in arcmin$^2$.  In both panels, circles refer to \XMM\, data,
  squares to {\it Chandra} data and triangles refer to {\it
    Beppo-SAX} data.}
\label{fig:t2d-EI}
\end{figure}

In Fig. \ref{fig:t2d-EI}, we plot the averaged profiles for $T$ and
$EI/Area$, respectively in panels (a) and (b). Circles refer to \XMM\,
data, squares to {\it Chandra} data and triangles refer to {\it
  Beppo-SAX} data.  The values for $T$ and $EI/Area$ are also reported
in Table \ref{tab:t2d-ei}. The emission integral is reported and
plotted in {\sc{xspec}} units, $[{{10^{-14}} / {4\pi
    d^2_{ang}(1+z)^2}}]EI $ where $d_{ang}$ is the angular distance of
the source in cm, $z$ is the redshift and $EI$ is in cm$^{-3}$.
$Area$ is the area of the ring in arcmin$^2$.  Note that error bars
for the $EI$ are quite small, especially for the \XMM\, data thanks to
its high effective area which allows a very precise measure of the
emission integral up to $\sim 100$ kpc. The $EI/Area$ profiles from
the three different data sets, match each other in the common ranges.
On the contrary, Fig.  \ref{fig:t2d-EI}(a) shows a systematic
difference between the three temperature profiles.  The discrepancy
between \XMM\, and {\it Beppo-SAX} is probably related to the use of
different energy bands.  For the observation of the Virgo cluster
which has a temperature of $2.5-3$ keV, the \XMM\, energy range
($0.4-4$ keV) is more suitable than the {\it Beppo-SAX} energy range
($2-10$ keV) and, consequently, \XMM\, estimations are probably more
reliable.  For what concerns the discrepancy between \XMM\, and {\it
  Chandra} temperature profiles, the differences are of the order of a
few percent and well within the cross-calibration uncertainties
between {\it Chandra} and \XMM\,.  To make full use of the combined
{\it Chandra}, \XMM\, and {\it Beppo-SAX} data and at the same time
avoid unphysical jumps in deprojected quantities when moving from one
dataset to the next, we decided to shift, through a scale
renormalization, the {\it Chandra} and the {\it Beppo-SAX} datasets.
\XMM\, was chosen as reference dataset because of its higher
statistics.  The scale factors have been derived by imposing that the
three temperature profiles match in the common ranges.  We find a
renormalization factor $\varkappa=1.03$ for the {\it Chandra}
temperature profile, and $\varkappa=1.10$ for the {\it Beppo-SAX}
temperature profile.

\begin{figure}
\epsscale{0.8}
\plotone{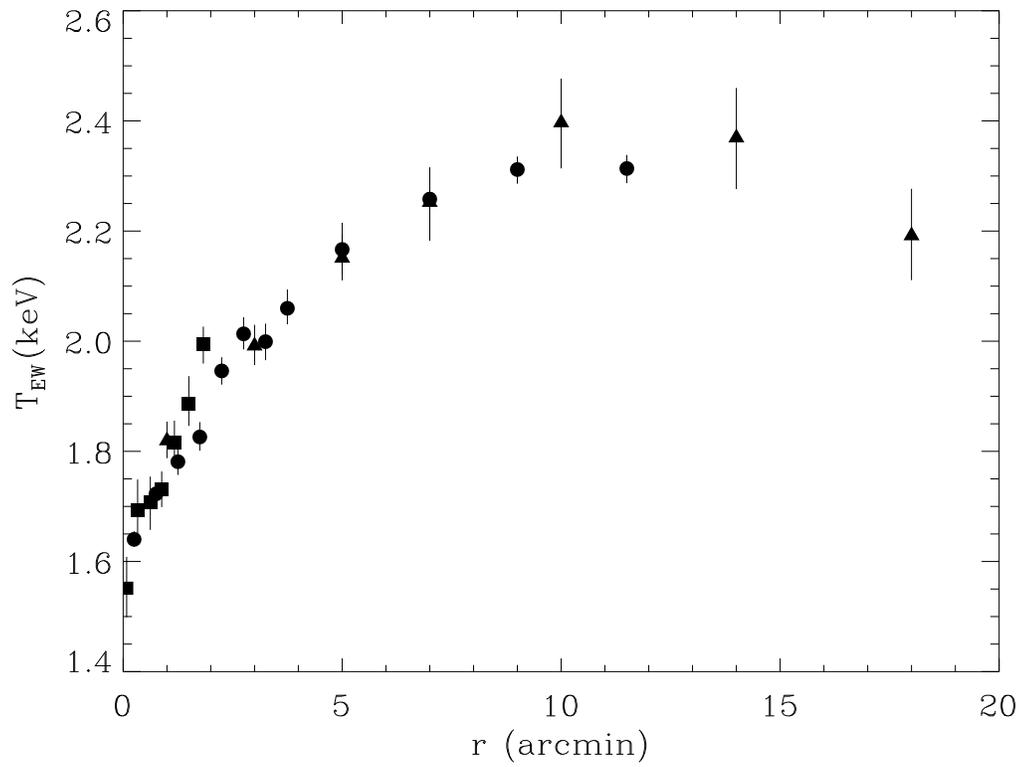}     
\caption
{Emission-weighted temperature profile for M87 obtained from the
  spectral analysis described in \S \protect\ref{sec:sp2d}, with the
  {\it Chandra} and {\it Beppo-SAX} data sets renormalized in order to
  match the \XMM\, profile. Symbols are the same as in Fig.
  \protect\ref{fig:t2d-EI}.}
\label{fig:tew3r}
\end{figure}

The temperature profiles, corrected for the scale factor, are plotted
in Fig.  \ref{fig:tew3r}.  The final joint data set used for the
analysis is given by (i) the {\it Chandra} data in the $0'-1'$ range
(4 points); (ii) the \XMM\, data in the $1'-8'$ range (8 points);
(iii) the {\it Beppo-SAX} data in the $8'-12'$ range (3 points).

\section{Spectral deprojection}
\label{sec:depro}

The deprojection technique has become very popular to investigate the
intracluster medium properties
\citep{ES02,Ett02,matsu,Allen96,pizzo02}.  Under the assumption of
spherical symmetry, the different 3-D quantities describing the ICM
are derived from the 2-D projected ones, starting from the outer shell
and moving inwards following an onion-peeling technique.

Among the different prescriptions available for the deprojection, we
decided to adopt the {\it spectral deprojection} introduced by
\citet{Ett02} \citep[see also][]{EDM02}. 
The physical quantities to be
deprojected are those obtained from the spectral analysis described in
the previous Section.  Each 3-D variable $f_{shell}$ is related to the
projected one $F_{ring}$ according to the relation

\begin{equation}
f_{shell} = \left( V^T\right)^{-1} \# F_{ring}, 
\label{eq:depro}
\end{equation}

\noindent
where $\left(V^T\right)^{-1}$ is the inverse of the transposed matrix
$V$ whose elements $V_{ij}$ are the volumes of the $i$-$th$ shell
projected on the $j$-$th$ ring.  The detailed evaluation of this
matrix can be found in \citet{Kriss83}. By replacing $F_{ring}$ with
(i) the emission-weighted measured temperature $T_{ring}L_{ring}$;
(ii) the ring luminosity $L_{ring}$ and (iii) $(EI/0.82)^{1/2},$ we
can derive respectively $\varepsilon T_{shell}$, the emissivity
$\varepsilon$ and the electron density $n_e$, where the relation $EI =
\int n_e n_p dV = 0.82 \int n_e^2 dV $ has been used. The main
advantage in using this technique is that deriving the electron
density $n_e$ from $EI$ is straightforward without any assumption on
its functional shape.  Moreover, since our measurements of the $EI$
are very accurate, an immediate and precise determination of the
electron density profile can be obtained.  It is worth noting that eq.
\eqn{eq:depro} is derived using the onion--peeling procedure where the
contribution to the emission in each shell is obtained from the
projected quantity by subtracting off the emission contribution of the
outer shells starting from the edge of the cluster and moving inwards.
In addition to this basic prescription, a correction factor accounting
for the cluster emission beyond the maximum radius $R_{max}$ to
infinity must be included.  The procedure to evaluate this correction
factor is presented in Appendix \ref{appen}.  In practice, in eq.
\eqn{eq:depro} the 2-D variable to be deprojected ($F_{ring}$) is
replaced by an effective one $F^{eff}_{ring}$ (see eq.
\ref{eq:feff}).

From now on, in order to avoid confusion, we will use $T$ for the 3-D
deprojected temperature and $T_{EW}$ for the 2-D emission-weighted
temperature.

\section{Deprojecting M87}
\label{sec:m87}

\subsection{Deprojected profiles}
\label{sub:depro}

\begin{figure}
\epsscale{0.65}
\plotone{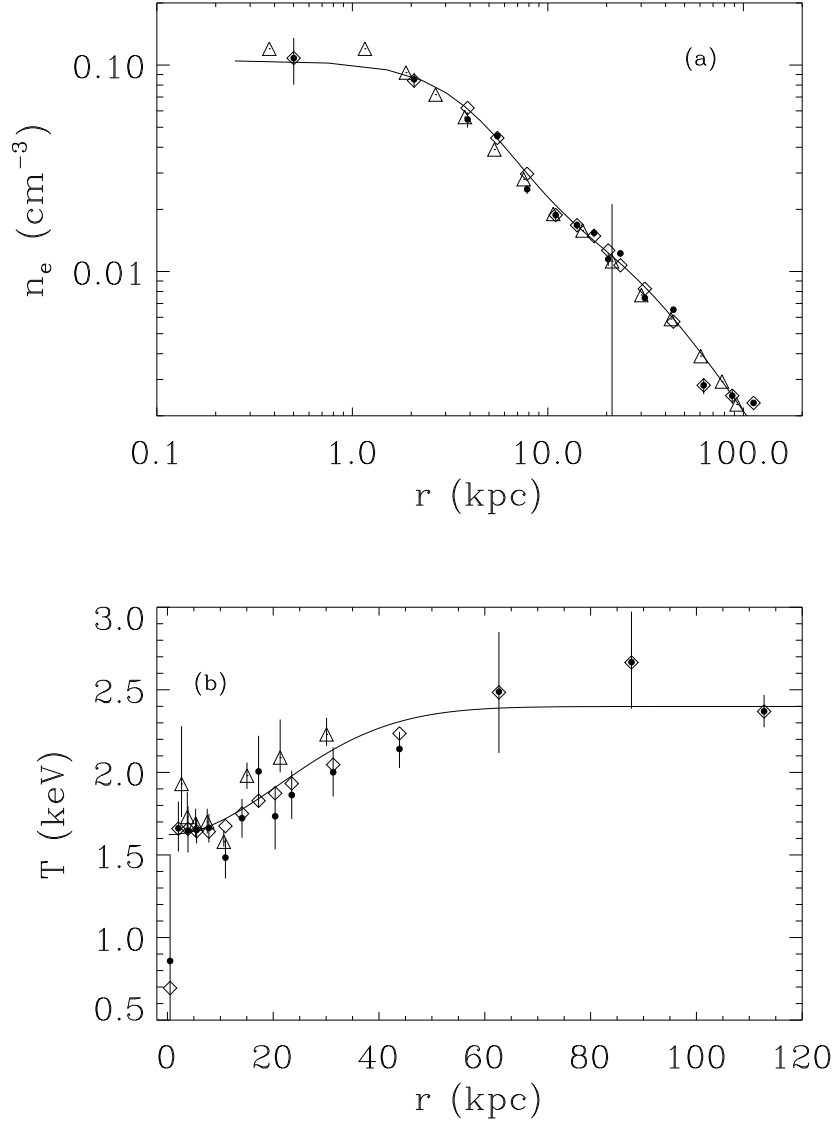}    
\caption
{ (a) Electron density derived with the deprojection method (filled
  circles).  The open diamonds represent the $n_e$ profile after the
  smoothing operation. The solid line is the best fit profile where a
  2-$\beta$ model (see eq.  \protect\ref{eq:2beta}) has been used. (b)
  Deprojected temperature profile versus radius (filled circles).  The
  open diamonds represent the $T$ profile after the smoothing
  operation.  The solid line is the best fit where the expression
  given in eq.  \protect\eqn{eq:tfit} has been used.  Error bars in
  both the panels have been obtained by 1000 Monte Carlo simulations
  on initial $EI$ and $T_{EW}$. The details on the smoothing operation
  and on the determination of the error bars are discussed in \S
  \protect\ref{sub:depro}. The triangles are the deprojected
  density and temperature profiles derived by \citet{matsu}.}
\label{fig:nel_tempfit}
\end{figure}

Following the prescription described in the previous Section, we
derive $\varepsilon$, $n_e$ and $T$ profiles for M87. All the
extracted values are reported in Table \ref{tab:tneps}.  In Fig.
\ref{fig:nel_tempfit} we plot (filled circles) the deprojected
electron density $n_e$ and temperature $T$, respectively in panel (a)
and (b).  For comparison, in Fig.  \ref{fig:nel_tempfit} we overplot
(open triangles) the deprojected profiles obtained by \citet{matsu}
who used a different choice of regions and a somewhat different
technique for spectral deprojection. The \citet{matsu} results plotted
here are those obtained by fitting the MOS data with a 2T model.  Our
profiles are in reasonable agreement with those derived by
\citet{matsu}.

Particular attention must be paid in the computation of the error bars
for $T$ and $n_e$.  In evaluating errors, we want to consider that,
even if the deprojection technique assumes that the spherical symmetry
condition is fulfilled, the dispersion of the $EI$ and $T_{EW}$
measurements along each ring around the averaged value is often
significantly larger than the error of each measure.  In particular,
this occurs for the \XMM\, data, since the \XMM\, large effective area
allows a very good statistics for M87 and provides $EI$ measurements
with very small error bars.  The scatter of the data around the
averaged value of the ring is a measure of the data displacement from
the spherical symmetry.  In order to account for this displacement in
the final error evaluation, we decided to assign to the $EI$ and
$T_{EW}$ of each region, an error ($\sigma_{EI}$ and $\sigma_{T_{EW}}$
respectively) which is the linear sum of two different contributions.
The first contribution is simply the error derived from the spectral
fit with the 1T/2T model. The second contribution is given by the
dispersion of the measurements along the ring around the averaged
value.  Error bars for the deprojected quantities $n_e$ and $T$
reported in Fig.  \ref{fig:nel_tempfit} have been obtained running
1000 Monte-Carlo simulations, sampling the $EI$ and $T_{EW}$ of each
region around their mean value assuming Gaussian distributions for the
errors $\sigma_{EI}$ and $\sigma_{T_{EW}}$.

The profiles plotted in Fig. \ref{fig:nel_tempfit} will be used as
starting points to derive some other quantities (such as mass and
conductivity).  In most cases, gradients of $n_e$ and $T$ are
involved.  Some smoothing procedure will be required to manage these
derivatives.  Consequently, we apply a smoothing algorithm which
replaces each point with the average value obtained using boxes of 3
points, i.e.: $V_i=\left(V_{i-1}+V_i+V_{i+1}\right)/3$.  In smoothing
the temperature profile, we excluded from the smoothing procedure the
two last ({\it Beppo-SAX}) points, in order to preserve the final
decreasing behavior of the $T$ profile. The final temperature profile
is obtained by applying the smoothing procedure twice: firstly, we
smooth the starting deprojected profile and then the obtained values
are smoothed again.  For the density profile, we applied the smoothing
procedure separately to the first 6 points and the others in order to
preserve the 2-$\beta$ behavior. Again the two last points have been
excluded from the smoothing operation and the smoothing has been
applied twice.  The open diamonds in Fig.  \ref{fig:nel_tempfit}
represent the $n_e$ and $T$ profile after the smoothing operation.  An
alternative solution to smoothing is provided by the use of analytical
functions fitting the profiles.  The solid lines in the figures are
the best fits to the data.  For the electron density profile we use
the fitting function:

\begin{equation}
n_e(r)= {{n_1} \over {\left[1+\left({r \over  {r_1}}\right)^2\right]^{\alpha_1}}} +
 {{n_2} \over {\left[1+\left({r \over  {r_2}}\right)^2\right]^{\alpha_2}}},
\label{eq:2beta}
\end{equation}

\noindent
which corresponds to a 2-$\beta$ model.  For the temperature profile
we find that the function

\begin{equation}
T(r)=T_0 - T_1 \exp\left( - {{r^2}\over{2\sigma_T^2}}\right)
\label{eq:tfit}
\end{equation}

\noindent
provides a good description of the data.  For the temperature profile,
we find the following best-fit values: $T_0= 2.399 \pm 0.090$ keV,
$T_1=0.776 \pm 0.097$ keV, $\sigma_T=3.887' \pm 0.731'$. For what
concerns the electron density, we find $r_1 =2.68' \pm 0.54'$,
$\alpha_1 = 0.71\pm 0.06$, $n_1=0.033 \pm 0.01$ cm$^{-3}$, $r_2=3.73'
\pm 9.65' $, $\alpha_2= 20.19 \pm 104.84$ and $n_2 =0.069 \pm 0.010$
cm$^{-3}$.  The inferred best fit values have large statistical
errors. This is due to the large number of free parameters adopted.
So, we decided to fix two parameters, namely, the core radius and the
slope of the second component; we fix the slope at large radii
$\alpha_2$ to the value obtained from the RASS measurements
\citep{bohr94} setting $\alpha_2=0.705$ ($\beta=0.47$).  For the
radius $r_2$ we decided to use the value of the $\sigma_T$ inferred
from the best fit of the temperature profile, which defines the scale
radius for the rise of the temperature.  Having fixed the values of
$r_2$ and $\alpha_2$, we find: $\alpha_1=1.518 \pm 0.317 $, $n_1=0.089
\pm 0.011$ cm$^{-3}$, $r_1 = 0.834' \pm 0.175'$ and $n_2=0.019 \pm
0.002$ cm$^{-3}$.  Note that the inferred value of $r_1$ ($\sim 5 $
kpc) roughly corresponds to the AGN jet extension
\citep[e.g.][]{DiM02,Young}.

\subsection{Sector deprojection}
\label{sub:slices}

The deprojection method is based on the assumption of spherical
symmetry. As discussed in \S \ref{sec:sp2d}, the hot component in
those regions which are described by the 2 temperature (2T) model
behaves very similarly to the gas in the regions described by a single
temperature (1T) model.  However, we should like to verify if any
correlation of the hot component with the radio emission exists,
invalidating our assumption of spherical symmetry.  To this aim, we
divided the cluster in sectors and deprojected separately each sector.
Again we consider only the hot component for the 2T regions. The
sectors have been chosen according to the radio emission regions: the
$[30^\circ -120^\circ]$, $[210^\circ-270^\circ]$ and
$[330^\circ-360^\circ]$ sectors have been cumulated together to form
the non-radio sector.  Furthermore, we analyzed separately each of the
following sectors: $[0^\circ-30^\circ]$, $[120^\circ-150^\circ]$,
$[270^\circ-300^\circ]$, $[300^\circ-330^\circ]$ which correspond or
are close to the radio emission arms.

\begin{figure}
\epsscale{0.8}
\plotone{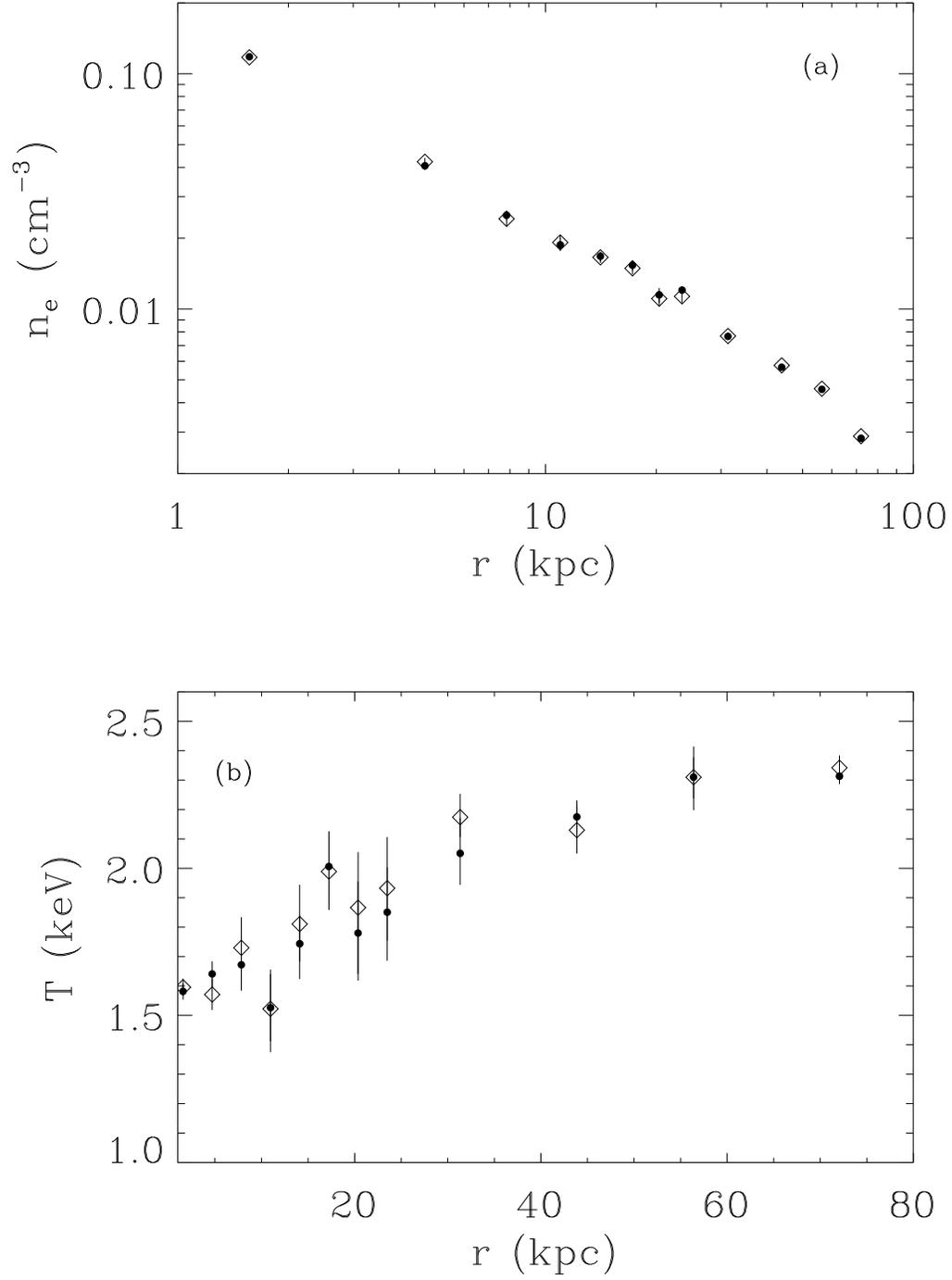}     
\caption
{ Deprojected density profile (a) and deprojected temperature profile
  (b) versus the radius for the whole cluster (filled dots) and for
  the non-radio regions (open diamonds).  No significant differences
  are evident. }
\label{fig:ne_tslice}
\end{figure}

The {\it Chandra} data are not suitable to perform such a study,
because the statistics is not very high. In this case, possible
azimuthal variations can be hidden by errors.  Therefore, in order to
verify the assumption of spherical symmetry, we consider only
the \XMM\, data (circles in Fig. \ref{fig:t2d-EI}).
Nevertheless, also with \XMM\, data, it is quite difficult to use
small sectors to perform sector-by-sector comparisons since their
statistics is not very high.  A significant comparison can be made
between the whole cluster and the non-radio selected sector.  In Fig.
\ref{fig:ne_tslice} (in panel (a) and (b) respectively) we compare the
$n_e$ and $T$ profiles for the whole cluster (filled dots) with the
non-radio sector (open diamonds).  In general, no significant
differences are evident.  The profiles of the electron density are
almost identical whereas there is a slight tendency of the temperature
calculated on the whole cluster to be smaller than the temperature of
the non-radio sector.  The difference is due to the fact that in the
non-radio sector only 1T regions contribute, while in the whole
cluster profile also the contribution of the hot component temperature
of the regions described by a 2T model is accounted. This temperature
is slightly smaller than the overall temperature and produces a mild
decrease of the whole cluster temperature profile with respect to the
non-radio sector temperature profiles. However, differences are well
within 1$\sigma$; thus, excluding the radio emission sectors (or the
regions described by the 2T model) does not affect significantly the
results and no evident azimuthal asymmetry can be highlighted. We can
conclude that the spherical symmetry assumption, which is an important
condition for our analysis, is tenable.  It is also worth noting that,
in order to derive most of the other physical quantities (mass,
conduction, entropy, etc.) a smoothing operation on $T$ and $n_e$ is
necessary, so that the small differences reported above would not
anyhow affect our results.

\subsection{The gravitational mass for M87}
\label{sub:mass}

\begin{figure}
\epsscale{1.}
\plotone{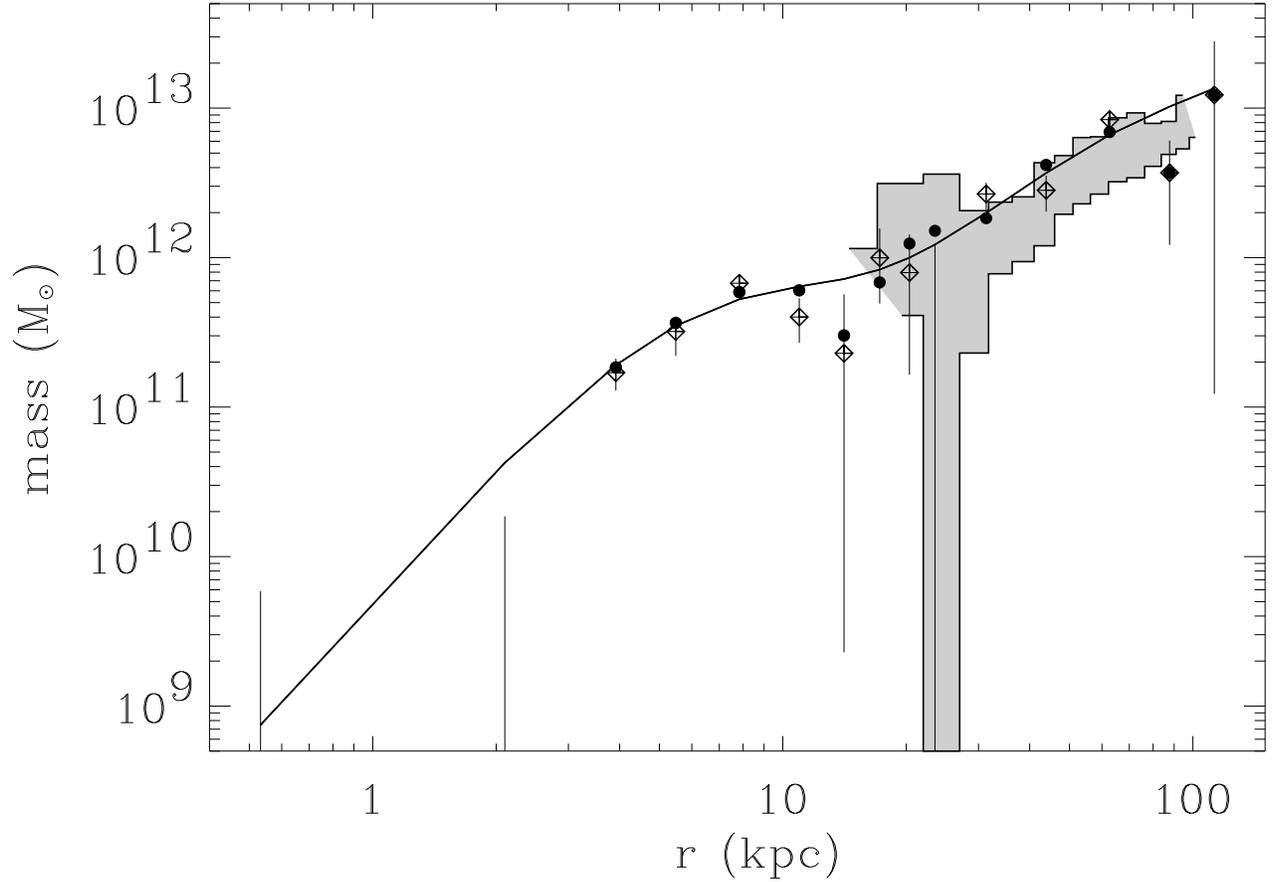}   
\caption
{Gravitational mass derived through the hydrostatic equilibrium
  equation.  The filled circles refer to the profile inferred by
  smoothing the temperature and the density profiles.  The open
  diamonds plot the gravitational mass derived without any smoothing
  operation on $n_e$ and $T$, errors have been derived using the
  standard Monte Carlo technique. Three of these points have values
  near to zero and only the upper limit of their error bar has been
  plotted here. The solid curve is the analytical mass obtained using
  the best fit profiles for $n_e$ and $T$ (eqs.
  \protect\ref{eq:2beta} and \protect\ref{eq:tfit}).  For comparison,
  the grey-shaded regions report the gravitational mass derived by
  \citet{NB95} using ROSAT--PSPC data.}
\label{fig:mass}
\end{figure}

Once the basic quantities $n_e$ and $T$ are obtained through the
deprojection, under suitable assumptions, other related quantities
describing the ICM can be derived.  One of the most important is the
gravitational mass. Supposing that the gas is in hydrostatic
equilibrium within the potential well of the dark matter, the
gravitational mass $M$ within a radius $r$ can be derived via the
hydrostatic equilibrium equation:

\begin{equation}
M( < r) = - {{kTr} \over {G \mu m_p}} \left[ {{d \ln{T} } \over {d \ln{r}}}
 + {{d\ln{n_e}} \over {d \ln{r}}} \right] 
\label{eq:hee}
\end{equation}

\noindent
where $\mu =0.6$ is the mean molecular weight, $G$ the gravitational
constant and $m_p$ the proton mass.  Even if eq.  \eqn{eq:hee}
provides a direct method to derive the gravitational mass, it is a
differential equation and the temperature and the electron density are
involved through their gradients.  Irregular features in the profile
induce jumps on the evaluated mass.  A classic solution consists in
smoothing the data and replacing the $n_e$ and the $T$ profiles with
their smoothed counterparts plotted in Fig.  \ref{fig:nel_tempfit}
(open diamonds).  As far as the temperature is concerned, a smoothing
procedure is viable since errors are rather large and the general
shape of the profile is quite smooth.  Correspondingly, the smoothed
profile is compatible (always within 1$\sigma$) with the original one.
On the contrary, for the electron density where errors are small, the
smoothing procedure could hide some features which are physical.  In
order to assess whether the smoothing affects results, we compare the
gravitational mass obtained using the temperature and the density
smoothed profiles with the gravitational mass derived using the {\it
  unsmoothed} profiles where errors have been evaluated using the
standard Monte Carlo technique.  As we show in Fig. \ref{fig:mass},
the smoothed profile (filled circles) agrees with the
non-smoothed one (open diamonds) within $1-2 \sigma$ and no
significant difference can be highlighted. It is worth noting that the
mass derived without smoothing provides three mass values (at $r \sim
0.5, \sim 2$ and $\sim 25$ kpc) which are negative and compatible with
zero: $ M=-0.5^{+1.1}_{-1.4} \times 10^{10}{\rm M_\odot}$,
$M=-2.7^{+9.6}_{-4.7} \times 10^{10}{\rm M_\odot}$ and
$M=-0.22^{+1.45}_{-1.67} \times 10^{12} {\rm M_\odot}$. For these
points, in Fig. \ref{fig:mass}, we show only the upper limit of the
error bar.

Alternatively to the smoothing procedure, the analytical expressions
for $n_e$ and $T$ (eqs. \ref{eq:2beta} and \ref{eq:tfit}) can be used.
The curve in Fig. \ref{fig:mass} is the analytical mass obtained
using these two best fit profiles. This mass profile has a plateaux at
a radius of about $10-15$ kpc ($\sim 2$ arcmin). This flattening
behavior is the consequence of the flattening of the 2-$\beta$
profile, at the same radius, which could correspond to the edge of the
central cD. The analytical gravitational mass is very similar to the
mass obtained both with the smoothing procedure and with the
unsmoothed data. The differences with respect to the latter curve are
limited to a few points ($\sim 15, \sim 25$ and $\sim 90$ kpc). For
these points, some ``holes'' appear in the shape of the non-analytical
profiles.  It is worth noting that the ``hole'' in a (integrated) mass
profile is not physical.  It may indicate that in this region the
hydrostatic equilibrium hypothesis breaks down \citep[e.g.][]{pizzo02}
and that an outflow providing additional pressure to support gravity
occurs there (see next Section for equations and further details).  In
any case, the error bars of all these points are large enough to make
``holes'' compatible with the analytical profile and no strong
evidence is present to claim that an outflow is present.

For comparison, the grey-shaded regions in Fig. \ref{fig:mass} report
the gravitational mass derived by \citet{NB95} using ROSAT--PSPC data.
Our profile, in the common radial regions is in agreement. It is
interesting to note that the point at $r \sim 20$ kpc has a very large
error bar in both the estimations.

\section{Cooling, conduction and heating}
\label{sec:heat}

As previously outlined, finding a heating mechanism able to balance
the cooling is not an easy task.  This Section will be
devoted to the inspection of some heating sources using the M87 data
set.

The heating contribution required to balance the radiative cooling can
be estimated starting from the thermodynamic equations describing a
spherically symmetric cluster:

\begin{eqnarray}
\nonumber
&&\frac 1 {r^2} {d \over {dr}}(\rho v r^2) = 0 \\
&&\rho v  {dv \over {dr}} + {d\over {dr}} \left({{\rho kT}\over {\mu m_p}}\right)  
 + {{GM} \over r^2}\rho = 0 
\label{eq:thermo}\\
\nonumber 
&&\frac 1 {r^2} {d \over {dr}} \left[ r^2\rho v \left( \frac12 v^2 + 
\frac 52 {{kT} \over {\mu m_p}} + \phi \right) \right] = -\varepsilon 
+ \varepsilon_{cond} + \cal{H}
\end{eqnarray}

\noindent
where $\phi$ is the gravitational potential, $M$ the gravitational
mass within $r$, $T$ the gas temperature, $\varepsilon$ the emissivity
and $\rho$ the gas mass density which is related to the electron
density according to $n_p=0.82 n_e= \rho/(2.21\mu m_p)$ ($\mu=0.6$ is
the mean molecular weight). The equations take into account the
possible presence of an inflow or outflow and the flow velocity $v$ is
taken positive outwards.  

The three equations \eqn{eq:thermo} are respectively the mass,
momentum and energy conservation equations. They have been derived
\citep[see][]{Sarazin} assuming a steady state; the second equation
reduces to hydrostatic equilibrium, eq.  \eqn{eq:hee}, for $v=0$.  We
include in the {\it right-hand-side} of the last equation, the
radiative cooling $\varepsilon$, and a heat contribution which
includes two parts: the thermal conduction $\varepsilon_{cond}$ which
will be widely discussed in the next paragraph, and a generic
extra-heating term $\cal{H}$, which will be studied in detail in
Section \ref{sub:rbmod}.

\subsection{Radiative cooling and conduction}
\label{sub:cond}

One obvious heating source is the thermal conduction which operates
when temperature gradients occur, and which can have a relatively
large efficiency (a fraction of the Spitzer conduction) even in
presence of magnetic fields \citep{gruz02,NM01}.

Neglecting any extra-heating source ($ \cal{H}$ $= 0 $ in eq.
\ref{eq:thermo}), and under the assumption of a spherical, steady
state, isobaric cooling flow, the last equation of \eqn{eq:thermo} can
be rewritten in the form:

\begin{equation}
- {{\dot M} \over {4 \pi r^2}}{d \over {dr}} \left({{5kT} \over {2 \mu m_p}} \right)
= - \varepsilon + \varepsilon_{cond} \, .
\label{eq:cond}
\end{equation}
\noindent
where 
$\varepsilon$ is the emissivity, $\mu =0.6$ is the mean molecular
weight and $\dot M$ is the usual mass deposition rate of the cooling
flow. 

The heating due to thermal conduction $\varepsilon_{cond}$ is given
by:

\begin{equation}
\varepsilon_{cond} = 
\frac 1{r^2} {d \over {dr}} \left( r^2  \kappa {{dT} \over {dr}}\right) , 
\label{eq:epscond}
\end{equation}

\noindent
where $\kappa$ is the conductivity. For a highly ionized plasma,
$\kappa$ is given by the \citet{spitzer} formula:

\begin{equation}
\kappa=\kappa_S = {{1.84 \times 10^{-5} \left(T/{\rm \gr}\right)^{5/2} } \over 
{\ln{\Lambda}}} 
{\rm \, erg \, cm^{-1}\, s^{-1}\, \gr^{-1}} \, ,
\label{eq:spitzer}
\end{equation}

\noindent
where $\ln{\Lambda} \sim 40$ is the usual Coulomb logarithm.

\begin{figure}
\epsscale{1.}
\plotone{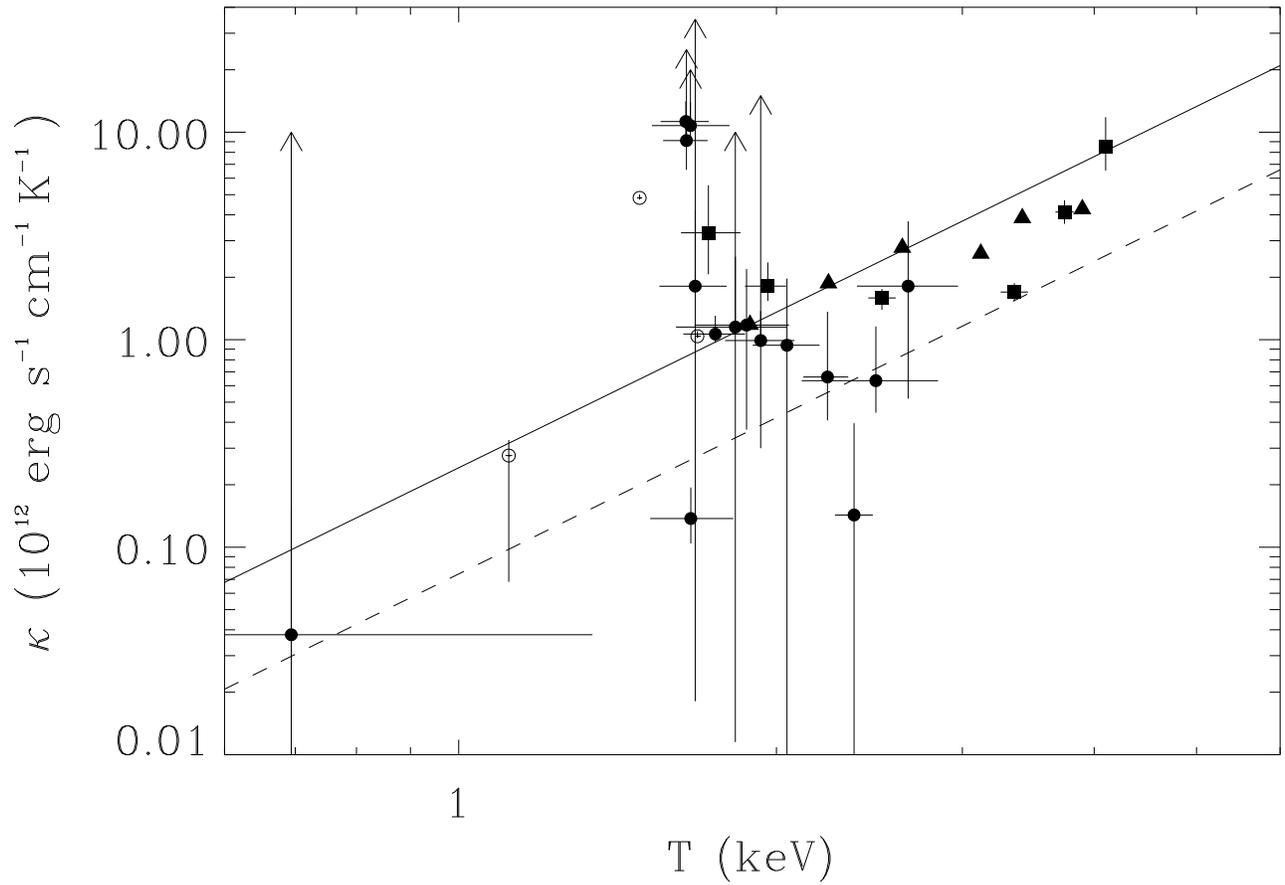}      
\caption
{The conductivity coefficient required for conduction to balance
  radiation losses, for M87 (close circles). The points which
  have an error bar which extents to infinity have been plotted with
  an arrow.  The solid line is the Spitzer conductivity and the dashed
  line is one third of the Spitzer conductivity. The open circles
  refer to results for M87 derived from \citet{VF04}. The triangles
  and the squares refer to \citet{Voigt} data for A2199 derived with
  two different prescriptions (see the text for details).}
\label{fig:cond}
\end{figure}

Starting from these equations we can derive the conductivity $\kappa$
required in M87 to stop the cooling flow ($\dot M = 0$).  The inferred
values for $\kappa$ are plotted in Fig. \ref{fig:cond} (filled
circles).  We plot $\kappa$ as a function of the temperature in order
to compare our results with \citet{Voigt} and \citet{VF04} who have
performed a similar calculation on a set of clusters. We recall that
the temperature grows with the radius.  Hence, the behavior of
$\kappa$ as a function of the temperature is similar to the behavior
of the profile of $\kappa$ as a function of the radius. From Fig.
\ref{fig:cond} we can see that the required conductivity has large
values for small temperatures (i.e. in the central part of the
cluster) and becomes smaller when the temperature increases, i.e.
moving towards the outskirts of the cluster. Note that the temperature
profile in the innermost regions of the cluster is consistent with
being constant.  Correspondingly, no conduction should be present and
the required conductivity is consistent with being as high as
infinity.  Hence, for these points, error bars will extent to
infinity.  In order to show in Fig.  \ref{fig:cond} that the error bar
for these points should extent to infinity, we plot their error bars
with an arrow.  The solid line in Fig.  \ref{fig:cond} represents
$\kappa_S$ given in eq.  \eqn{eq:spitzer} and dashed line corresponds
to 0.3$\kappa_S$ which could be the effective conductivity in presence
of turbulent magnetic fields \citep{gruz02}.  Fig.  \ref{fig:cond}
shows that in M87, the thermal conduction is able to balance radiative
cooling only in the outer part of the cluster.  For $r \simgt 10-20$
kpc the conductivity required for conduction to balance the cooling is
between 0.3$\kappa_S$ and $\kappa_S$.  In the inner $\sim 10-20$ kpc
in M87 the heating supplied by thermal conduction is not enough and an
extra-heating, whatever its source might be, is needed. The failure of
the thermal conduction in the core of the cluster is due to the fact
that in these regions the temperature profile flattens (see Fig.
\ref{fig:nel_tempfit}b) and the conduction decreases substantially.
At the same time, the innermost regions are those where the X-ray
emissivity is highest and which mostly require a heating source to
compensate the radiated energy.

In a recent paper, \citet{Voigt} determined the conductivity required
for the conduction to balance the radiation losses for A2199 whose
temperature is similar to M87 temperature ranging from $\sim 2 $ to $
\sim 5 $ keV.  For comparison, we plot in Fig. \ref{fig:cond} the
\citet{Voigt} data for A2199 derived by modeling the temperature
profile with two different prescriptions: a power law (triangles)
and a more complex functional form (squares) which flattens at
small and large radii \citep[see eq. (6) in ][]{Voigt}.  Because of
the larger distance of A2199 ($z=0.0309$) only a couple of points are
within the central 10 kpc.  Nevertheless, in agreement with our
results, they find that for these two central bins some extra-heating
is needed.  The quality of the M87 data set allows to highlight the
problem of conduction in the core and to analyze it in greater detail
than for A2199.  Clearly, M87 is a good object to test heating models.
In Fig.  \ref{fig:cond} we also plot (open circles) the $\kappa$
values obtained for M87 by \citet{VF04}. Their values are in
reasonable agreement with ours, although their analysis procedure is
quite different from ours. In \citet{VF04} only {\it Chandra} data are
included limiting the extension of the deprojected region to the inner
10 kpc and only the 1T {\sc xspec - mekal} model is used in fitting
spectra extracted from annuli.

\subsection{Heating for M87 and the ``effervescent''
  heating model }
\label{sub:rbmod}

In this Section we consider eqs. \eqn{eq:thermo} in their generic form
in order to determine the extra-heating $\cal{H}$ required to balance
the radiative cooling in presence of thermal conduction for M87.

Eqs. \eqn{eq:thermo} can be recasted in the form:

\begin{eqnarray}
\nonumber
&& v\rho r^2 = const = {{\dot M } \over {4 \pi}}\\
\label{eq:heating}
&& M = - {{r^2 v }\over {G}} {{dv} \over{dr}} -  {{kTr} \over {G \mu m_p}}
\left[ {{d \ln{T} } \over {d \ln{r}}}
 +  {{d\ln{n_e}} \over {d \ln{r}}} \right] \\
\nonumber 
&& {\cal{H}} =\varepsilon - \varepsilon_{cond}+\varepsilon^\star
\end{eqnarray}
\noindent
where we set:
\begin{equation}
\varepsilon^\star =   {{\rho v k T } \over {\mu m_p r}}
\left[ \frac 32 {{d\ln{T}} \over{d\ln{r}}} -  {{d\ln{\rho}} \over{d \ln{r}}} \right] \, .
\label{eq:epsstar}
\end{equation}

\noindent
This term includes the variation of the energy (per unit volume) due
to the outflow/inflow and the work (per unit volume) done by the
system during the outflow/inflow.  The mass flow rate $\dot M$ is
positive for an outflow and negative for an inflow.  The last equation
of \eqn{eq:heating} provides the heating $\cal{H}$ necessary to
balance the radiative cooling $\varepsilon$, in presence of thermal
conduction and steady outflow. The deprojected data $T$, $n_e$ (or
$\rho$), of M87 can be used to solve numerically eqs.
\eqn{eq:heating} for M87, deriving $M$, $v$ and $\cal{H}$, once we
have fixed the fraction $f$ of the Spitzer conductivity (eq.
\ref{eq:spitzer}) and some assumption has been made on $\dot M$.  We
fix an $f=0.3$ efficiency \citep[see][]{gruz02} and we set the outflow
mass rate $\dot M = 1.6{\rm M_\odot/yr}$.  This $\dot M$ value is
similar to the asymptotic value that RB02 obtained from their
simulations for the stable final state of the cluster.  We will
discuss further on different values for $\dot M$ and $f$.

As far as the velocity is concerned, for the assumed values of $f$ and
$\dot M$, $v$ is smaller than a few tens of km/s, for radii larger
than $\sim 1-2$ kpc.  Correspondingly, the term including the
velocity in the momentum conservation equation (the second equation of
\ref{eq:heating}) is significantly smaller than the total mass being
of the order of $10^5-10^7 \, {\rm M}_\odot$ versus the
$10^{10}-10^{12} \, {\rm M}_\odot$ of the total mass, so its
contribution is negligible.  Hence, we can state that the cluster is
almost in hydrostatic equilibrium and the mass estimates reported in
Fig. \ref{fig:mass}, where the term related to the outflow is
neglected, are not affected.  In order to have a significant
contribution from the velocity term and to alter substantially the
hydrostatic equilibrium, $\dot M$ values as large as several tens--few
hundreds of ${\rm M}_\odot/{\rm yr}$ are required.

For the considered values of $\dot M$, the quantity
$\varepsilon^\star$ in eq. \eqn{eq:epsstar} is negligible with respect
to the emissivity $\varepsilon$. Thus, the extra-heating $\cal{H}$
and the conduction term $\varepsilon_{cond}$ are completely used to
balance the radiative cooling.

\begin{figure}
\epsscale{1.}
\plotone{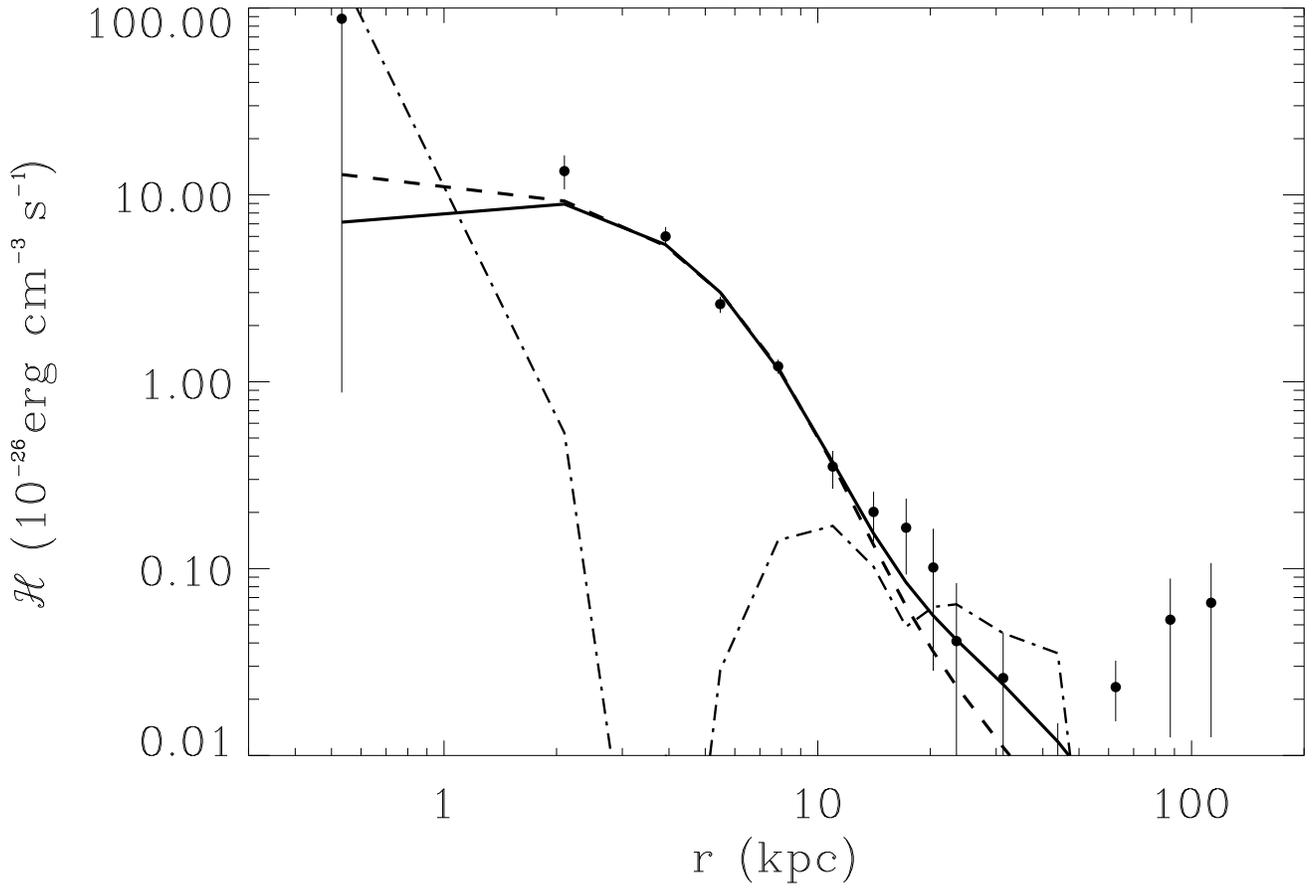}     
\caption
{The heating required (filled circles) to balance radiation
  losses in M87.  The dot-dashed line is the heating due to thermal
  conduction.  The solid line is the best fit obtained fitting
  the data set with the RB02 model.  The dashed line is the best
  fit obtained fixing $\gamma_b = 4/3$.}
\label{fig:heating}
\end{figure}

For what concerns $\cal{H}$, the heating required in M87 is plotted in
Fig. \ref{fig:heating} (filled circles).  As expected, most of
the heating is required in the central part of the cluster (say in the
inner $\sim 15-20$ kpc), where conduction is not efficient.  The
heating due to thermal conduction is plotted in Fig.
\ref{fig:heating} (dot-dashed line); in the central $10$ kpc,
where the temperature profile becomes flatter, the thermal conduction
drops to zero, apart from the innermost bin ($\sim 1$ kpc) where $T$
falls to very small values (see Fig. \ref{fig:nel_tempfit}b), with a
large error bar. The conduction in this bin is $\varepsilon_{cond}
=1.46^{+1.89}_{-1.53} \times 10^{-24} {\rm erg \, cm^{-3} s^{-1}}$,
and is in agreement with zero within 1$\sigma$.

The heating model developed in \citet[RB02 hereafter]{RB02} and \citet{beg01}
includes a mechanism for heat injection from the central AGN.  The
mechanism has been called ``effervescent heating''.  The radio source
is supposed to deposit some buoyant gas bubbles in the ICM, which do
not mix and rise through the ICM microscopically.  The bubbles should
expand doing work on the surrounding plasma and converting their
internal energy in heat.  The buoyant outflow contribution in the
energy conservation equation is accounted for in the
$\varepsilon^\star$ term, while $\cal{H}$ describes the heat injection
due to the adiabatic expansion of the bubbles.

According to the RB02 model, the heating $\cal{H}$ is a function of
the pressure (and its gradient) and can be expressed according to:

\begin{equation}
{\cal{H}} = - h(r) \left({p \over p_0}\right)^{(\gamma_b-1)/\gamma_b} \frac 1r 
{{d\ln{p}}\over {d\ln{r}}}, 
\label{eq:hbr}
\end{equation}

\noindent
where,  
\begin{equation}
h(r)= {L \over {4 \pi r^2}} \left( 1 - e^{-r/r_0} \right) q^{-1}
\label{eq:normf}
\end{equation}

\noindent
and

\begin{equation}
q=\int_{0}^{+\infty}{\left({p \over {p_0}}\right)^{(\gamma_b-1)/\gamma_b}
\frac 1r {{d\ln{p}}\over {d\ln{r}}} \left( 1 - e^{-r/r_0} \right) dr} \, ;
\label{eq:qint}
\end{equation}

\noindent
$p$ is the pressure, $p_0$ is the central pressure, $L$ the
time-averaged luminosity of the central source, $\gamma_b$ is the
adiabatic index of the buoyant gas and $r_0$ the inner heating cutoff
radius.  The term $ 1 - exp(-r/r_0)$ introduces an inner cutoff which
fixes the scale radius where the bubbles start rising buoyantly in the
ambient plasma.  $\cal{H}$ is normalized in such a way that, when
integrated over the whole cluster, the total injected power
corresponds to the time-averaged power output of the AGN.  $\cal{H}$
has been derived in eq. \eqn{eq:hbr} assuming a steady state for the
bubble flux.  In order to assess if this assumption is reasonable we
must compare the different timescales involved in the effervescent
heating mechanism.  We can suppose that the AGN is intermittent (RB02;
\citealp*{RB97}) and heats the ICM through a succession of outbursts.
During each outburst, the AGN injects a population of bubbles which
subsequently rise buoyantly.  During the ``off'' periods, the bubbles
continue their rise heating the cluster atmosphere. If the outbursts
follow each other on a timescale which is short with respect to the
rising timescale of the bubbles, then the flux of the bubbles reaches
a quasi steady state.  In fact, the ratio $t_{rise}/t_i$ between the
rise timescale $t_{rise}$ and the intermittence timescale $t_i$ gives
the number of populations injected within the ICM within a time
$t_{rise}$. The larger this ratio is, the larger the number of bubble
populations rising into the cluster atmosphere and the mechanism
approaches the steady state. The radio galaxies are likely to be
intermittent on a timescale as short as $t_i \sim 10^4-10^5$ yr (RB02;
\citealp*{RB97}). In the next section we will see that the risetime is
$t_{rise} \sim 10^8$ yr or even larger. The value of the ratio
$t_{rise}/t_{i}$ is $10^3-10^4$ or more; therefore the assumption of
steady state is tenable and the released heating may be treated in a
time-averaged sense.

RB02 also include a convection term in eq.  \eqn{eq:heating} which we
have neglected. The reason for this choice is twofold. First of all,
the convection must be limited to the innermost regions of the
cluster, in order to allow the presence of metallicity gradients in
cluster cores \citep{DM01}.  Most importantly, a negative gradient for
the entropy is a necessary condition for the onset of the convection.
In fact, the condition of instability:

\begin{equation}
{d \over dr}\left({p \over {\rho^\gamma}}\right) < 0 
\label{eq:convec}
\end{equation}

\begin{figure}
\epsscale{1.}
\plotone{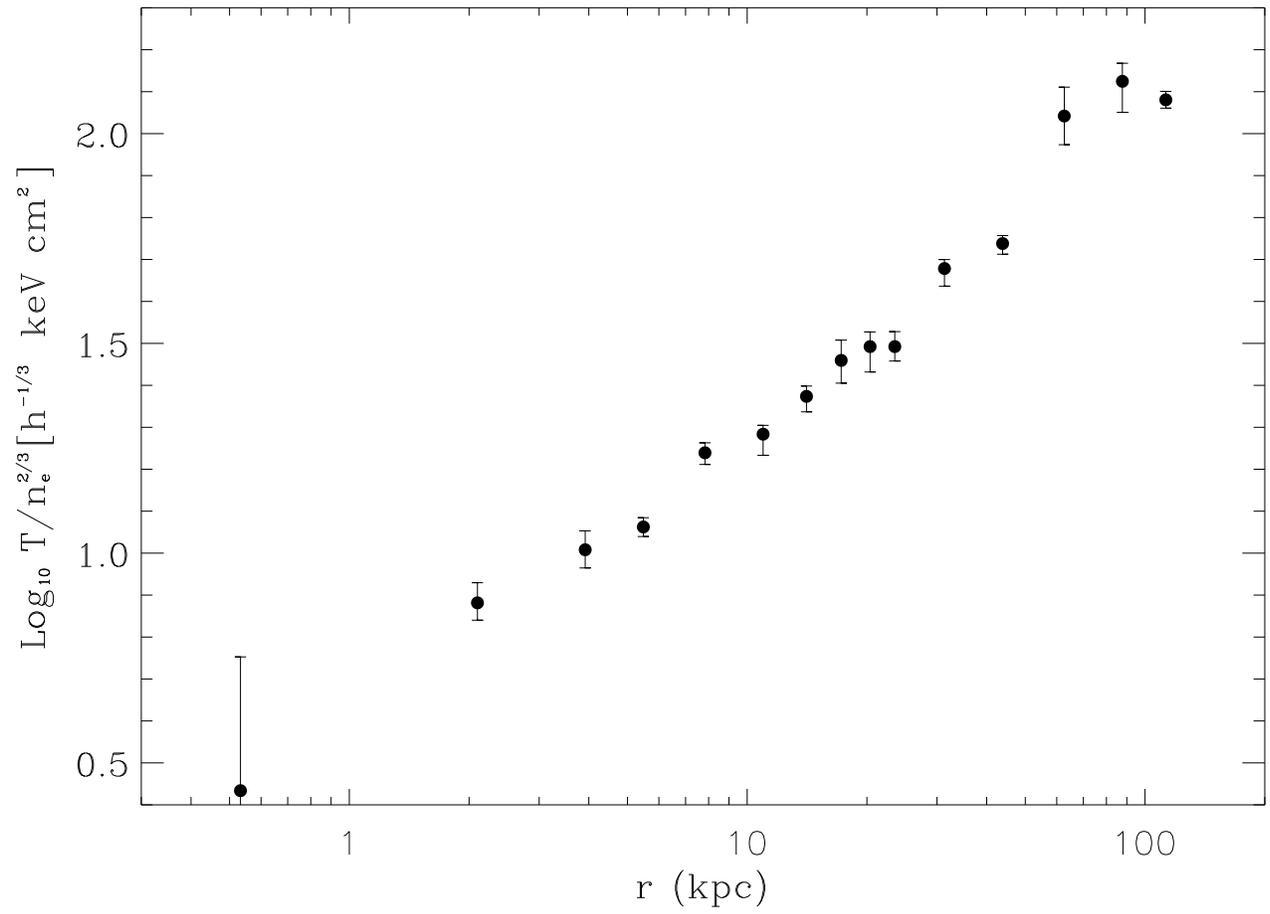}     
\caption
{Entropy profile for M87.}
\label{fig:entropy}
\end{figure}

\noindent
(where $\gamma$ is the ratio of the specific heats $c_p/c_v$ and has
the value $5/3$ for a highly ionized gas) must be fulfilled for the
convection to operate and it is equivalent to requiring that $\nabla
 (T/n^{2/3}) <0$.  This condition is not satisfied in M87 where we
verified that the entropy is an increasing function of the radius, as
we show in Fig. \ref{fig:entropy}.

We compare the values inferred for the extra-heating $\cal{H}$ term in
M87 reported in Fig.  \ref{fig:heating} (filled circles), with
the predictions from the RB02 model derived according to eq.
\eqn{eq:hbr}, in order to assess whether, for a reasonable choice of
the parameters, the heat flux required to balance the cooling is
compatible with the heat injected by the central AGN.
 
In order to determine the pressure and the pressure
gradient in eq. \eqn{eq:hbr}, we use the analytical expressions
\eqn{eq:2beta} and \eqn{eq:tfit} for $n_e$ and $T$ with the best-fit
parameters obtained by fitting the deprojected electron density and
temperature profiles.  We use eqs. \eqn{eq:hbr} - \eqn{eq:qint} as
fitting functions for the extra-heating, where $\gamma_b$, $r_0$ and
the total normalization $A$:

\begin{equation}
A={L\over{4\pi}}q^{-1}
\label{eq:normfit}
\end{equation}

\noindent
are the free parameters.  The solid line in Fig.  \ref{fig:heating} is
the derived best fit.  The model proposed by RB02 seems to provide a
fair description of the heating needed to balance the cooling flow in
the inner $\sim 15-20$ kpc of the cluster, where conduction is not
sufficient.  The line seems to follow adequately the behavior of the
data points.  Nevertheless, the derived fit values have large errors.
By fixing $\gamma_b = 4/3$ (which is the adiabatic index for
relativistic bubbles), we find $r_0=4.39 \pm 1.83$ kpc and $A = (8.35
\pm 1.65) \times 10^{-24}\, {\rm erg \, s^{-1} }$ (the dashed curve in
Fig.  \ref{fig:heating}).  By using eq.  \eqn{eq:normfit}, we can also
derive the central AGN luminosity $ L = 5.95 \times 10^{42} \,$ ${\rm
  erg \, s^{-1} }$ required to stop the cooling.
 
The three external points in Fig. \ref{fig:heating} seem to require an
additional extra heating, showing an excess with respect to the
general behavior of the data at large radii and with respect to the
best fit function shape.  This excess is related to the flattening of
the temperature profile in the external regions which dampens
conduction. However, it must be noted that the values of the heating
$\cal{H}$ required in these three bins are respectively:
$2.32^{+0.89}_{-0.80} \times 10^{-28} {\rm erg \, cm^{-3} s^{-1}}$,
$5.32^{+3.51}_{-4.07} \times 10^{-28}{\rm erg \, cm^{-3} s^{-1}}$ and
$6.56^{+4.14}_{-5.31} \times 10^{-28}{\rm erg \, cm^{-3} s^{-1}}$ and
are all in agreement with zero within $2-3 \sigma$.  Hence, the
evidence for the excess is not particularly strong.

Our estimated radius $r_0 \sim 4-5 $ kpc, is comparable to the
extension of the AGN jet as seen in the {\it Chandra} image of M87
\citep[e.g.][]{DiM02,Young}.  The fact that $r_0$ is of the same order
of magnitude of the jet is consistent with a scenario where the
effervescent bubbles are generated through the interaction of the
radio jet with the cluster atmosphere.  Understanding the precise
nature of such interaction will require considerable efforts both on
the observational and theoretical side.  Our aim here is simply to
note that our fit does not rule out the possibility that the bubbles
are generated through the interaction of the radio jet with the
cluster atmosphere, as would have been the case if, for example, the
fitting of the effervescent model to M87 had returned an $r_0$ value
10 times larger than the one actually measured.  The inferred
luminosity value is similar to the luminosity evaluated for the M87
AGN \citep[OEK hereafter]{owen}, $ \sim 3-4 \times 10^{42}{\rm erg \,
  s^{-1} } $ \citep[see also][]{DiM02}.  Slightly different values for
the luminosity could be inferred with different choices of $f$ and
$\dot M$ which of course provide different best fit parameters values
and different related luminosities.  When a larger $f$ is considered,
the higher contribution of the thermal conduction reduces the amount
of heating needed to balance the radiative cooling.  Setting the
conductivity to the Spitzer value ($f=1$) we find best fit values
which are similar to those inferred for $f=0.3$ and provide a slightly
lower luminosity ($\sim 2-3 \times 10^{42} {\rm erg \, s^{-1} } $).
On the contrary, when smaller values for $f$ are considered, some
additional heating is necessary.  For $f=0.1$ we derive an AGN
luminosity $\sim 1-2 \times 10^{43} {\rm \, erg\, s^{-1} }$, which is
somewhat larger than the OEK estimations.  However, for such small
efficiencies, the shape of the RB02 model no longer provides a good
description of the data, especially in the central regions.  Thus, if
the contribution of the thermal conduction is too small, the
``effervescent heating'' model is not suitable to describe the heating
necessary to balance the cooling flow.

Some variations are found also for different initial values of $\dot
M$.  We tried 16 and 0.16 M$_\odot$/yr corresponding to 10 and 0.1
times the original $\dot M$ value we considered.  As expected, for
large values of $\dot M$ the corresponding AGN luminosity is
significantly enhanced ($\simgt few 10^{43} \,{\rm erg\, s^{-1} }$)
since the central AGN must provide a larger quantity of energy to the
outflowing bubbles.  The variations when smaller $\dot M$ are
considered, are modest, slightly reducing the luminosity to $\sim 3-4
\times 10^{42} \, {\rm erg\, s^{-1} }$.

Note that it is not necessary for the AGN luminosity, obtained by
requiring that the ``effervescent heating'' model balances the cooling
flow to be exactly equal to the AGN luminosity derived from radio
observations. In fact, the required luminosity from the model should
be regarded as a time-averaged power of the AGN, as the AGN dynamical
times are smaller than the radiative cooling flow scale times.  One
should also keep in mind that only a fraction of the total power of
the AGN is used to quench the cooling flow and that the luminosity
required from the model can be significantly smaller than the real AGN
luminosity.  From our analysis, we can infer that the values of $L$
derived with different choices of $f$ and $\dot M$ are of the same
order of estimates by OEK and \citet{DiM02}.

While the luminosity is slightly affected for different choices of
$\dot M$ and $f$, $r_0$ variations are quite modest and the inferred
values of the scale radius are always of the order of $r_0 \sim 4-5$
kpc, which approximatively corresponds to the AGN jet extension.

\subsection{Discussion}
\label{sub:discussion}

Starting from the results inferred in the previous section, we can try
to draw a more general picture, using also informations coming from
radio observations of M87.

We can suppose that the buoyant bubbles are radio bubbles filled with
magnetic field and relativistic particles
\citep{GN73,ch00,ch01,BK01,ch02} responsible for the synchrotron
emission in M87.  As already outlined by OEK, the
radio structures are highly filamented.  This suggests that the
dimensions of the bubbles are small.  \citet{EH02} discussed the
dynamics of the rise of buoyant light bubbles within the cluster
atmosphere \citep[see also][]{ch01,Kai03}. The buoyant bubble rapidly
reaches a terminal velocity $v_b$. In the limit of small bubbles, $v_b$
can be estimated by balancing the buoyancy force with the ram pressure
(drag force) of the cluster gas.

The buoyancy force is 

\begin{equation}
\label{eq:fbuoy}
F_b = V g (\rho  - \rho_b)  \, = V g \rho \Delta \, ,
\end{equation}
\noindent
where $V = 4/3 \pi r_b^3$ is the volume of the bubble, $r_b$ is the
bubble radius, $g=G M(<r)/r^2$ is the local gravity acceleration
at the radius $r$ ($M$ is the gravitational mass within the radius
$r$); $\rho$ and $\rho_b$ are respectively the density of the ICM and
of the bubble. $\Delta =(\rho-\rho_b)/\rho$ is the density contrast.

The drag force for subsonic motion can be approximated by

\begin{equation}
\label{eq:drag}
F_{drag} = C \pi r_b^2 \rho v_b^2
\end{equation}
\noindent
with the drag coefficient $C\simeq 0.5$.

By equating eqs. \eqn{eq:fbuoy} and \eqn{eq:drag} the velocity of the
bubble can be determined.  \citet{EH02} derived $v_b$ under the
assumption that the density of the ICM is well described by an
isothermal $\beta$-model and the density contrast $\Delta \sim 1$. In
this case $v_b$ is a fraction ($\propto \sqrt{r_b/r_c}$; $r_c$ is the
core radius of the cluster) of the sound velocity.  Correspondingly,
for small bubbles, $v_b$ is subsonic which is consistent with the fact
that no shocks are detected in M87.  

\begin{figure}
\epsscale{1.0}
\plotone{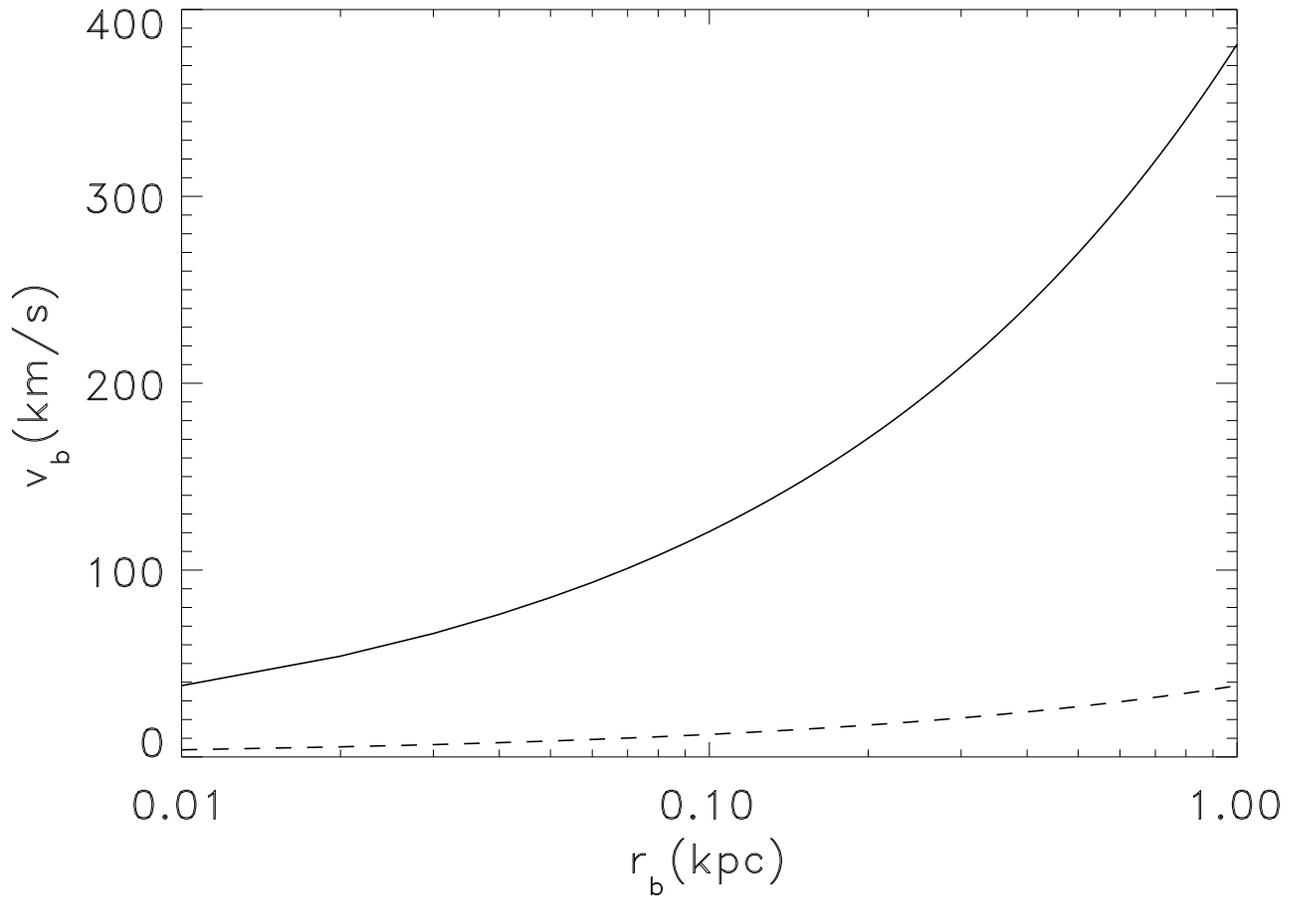}     
\caption
{ Rise velocity for the buoyant bubbles for density contrast between
  the bubble and the ambient ICM $\Delta =1$ (solid line) and
  $\Delta=0.1$ (dashed line). All the quantities have been evaluated
  at $r\sim 10$ kpc.}
\label{fig:velb}
\end{figure}

We derived the rise velocity of the bubbles in M87 at a radius $r \sim
10$ kpc, using the results from the deprojection for $M(<r)$ and
$\rho$.  The density contrast $\Delta$ can be inferred considering
that the bubbles filled with relativistic plasma are in pressure
equilibrium with the ICM.  Indeed, the pressure equilibrium between
radio bubbles and thermal plasma implies that $n_e kT \sim \beta
n_{er} \langle E_e\rangle$ (where $\beta$ is a factor of the order of
the unity which accounts for the contribution to the pressure of the
magnetic field which is almost in equipartition with the particle
energy).  $\langle E_e\rangle$ is the mean energy of the relativistic
particles and $n_{er}$ is the numerical density of the relativistic
particles. The relativistic particles mean energy is much larger than
that of the thermal electrons in the ICM, leading to a density
contrast $\Delta \sim 1$.  The values of $v_b$ for the bubble
radius $r_b$ varying in $[0.01-1.]$ kpc range are plotted in Fig.
\ref{fig:velb} and the motion is indeed subsonic.

It is worth considering that, in a recent analysis of these data 
\citep[M02 hereafter]{ms02}, we have found that, cospatially with the
radio lobes, there exists a thermal component with $T \sim {1\over2}
T_{ICM}$. This component is likely structured in blobs with typical
scales smaller than $\sim 100$ pc.  The filling factor of these blobs
has been estimated to be of the order of the percent.

Some informations about the filling factor of the radio bubbles can be
obtained using recent results from radio observation of M87.  Using
the standard minimum pressure analysis \citep[e.g.][]{Pacho70,bor,OO},
OEK evaluate the magnetic field in the lobes and in the filaments
visible in the radio map and evaluate the minimum pressure of the
magnetic field and relativistic particles.  They assume that the
proton-to-electron energy is $k=1$ and that $\zeta\phi=1$ where $\phi$
is the filling factor of the relativistic particles and of the
magnetic field and $\zeta$ is the ratio of the true magnetic pressure
(comprehensive of the tension) to the magnetic pressure $B^2/8\pi$.
For a tangled magnetic field $\zeta=1/3$.  The estimations for the
minimum pressure derived by OEK can be compared with the thermal
pressure that we can infer from our deprojected $T$ and $n_e$
profiles.  In agreement with OEK, we find that the pressure of the
relativistic plasma is smaller than the thermal pressure by about an
order of magnitude.  By keeping the condition $k=1$ and assuming
$\zeta=1$, we can derive the filling factor which reconciles the
minimum pressure $P_{min}$ with the thermal pressure of the plasma,
considering that the minimum pressure scales with $\phi$ according to:

\begin{equation}
{{P_{min}(\phi) } \over {P_{min}(\phi=1)}} =\left(\phi\right)^{-4/7} \, .
\end{equation}

The discrepancy of a factor $\sim 10$ between the two pressures can be
eliminated by assuming a filling factor $\sim 1 \%$. By using the
estimations of $P_{min}$ from OEK in different plasma of the radio
lobes and filaments, we infer filling factors of few $\%$. Assuming
$\zeta=1/3$ (tangled magnetic field) reduces $\phi$ by a factor $\sim
2$.  The filling factors of the radio bubbles and of the cold thermal
blobs are of the same order.

The survival of the cold thermal blobs in the hotter ICM requires that
thermal conduction be substantially suppressed; this may happen if
these blobs are tied to the radio bubbles and magnetic fields shield
them from collisions with ICM particles. Since the filling factors of
blobs and bubbles are similar and the density of the blobs is about
twice that of the ICM (M02), the mean density of each bubble+blob is
about that of the surrounding ICM.  Thus, assuming that the blobs are
tied to the bubbles and that they occupy similar volumes, their
density contrast with the surrounding ICM will be small. In Fig.
\ref{fig:velb} the dotted line plots the bubble rise velocity $v_b$
for a $\Delta=0.1$.  Starting from the rise velocity, it is
straightforward to derive the rise timescale of the bubbles.  For a
density $\Delta \sim 1$ and a bubble radius of $\sim 100$ pc, the
risetime is $t_{rise} \sim 10^8$ yr at $r \sim 10$ kpc and it becomes
even larger for smaller bubbles and smaller density contrasts, holding
$t_{rise} \propto {(r_b \Delta)}^{-1/2}$.  As outlined previously,
this value is significantly larger than the duty cycle timescale $t_i$
of intermittency of the AGN, so that the ratio $t_{rise}/t_i \simgt
10^3-10^4$ and the mechanism approaches the steady state.

It is worth noting that the above picture refers to radio bubbles and
cool X-ray blobs located in the lobes. However, the heating process
related to the rise of the bubbles must be isotropic throughout the
cluster in order to balance the cooling flow.  Under the assumption
that the radio galaxy is intermittent on a short timescale, the
mechanism is expected to heat isotropically the ICM since it is likely
that no direction is preferred for the AGN ejection; we also recall
that when the AGN is turned off the bubbles continue their rise within
the ICM heating it up; within this picture, by averaging on a cooling
time, the radio bubble populations are likely to be isotropically
distributed in the cluster. 

Nevertheless, the cool X--ray blobs are detected only in the lobes
regions.  So the picture emerging here is that of an AGN which injects
radio bubbles, in all the directions, through a succession of
outbursts.  In the lobes, the outburst is occurring at the present
time and, here, also the cool X-ray blobs are present. They are tied
to the radio bubbles and the magnetic fields shield them from
collisions with ICM particles. The thermal conduction is inhibited
allowing the cool blobs to survive in a hotter ambient medium.  In the
halo, where the radio bubbles have been injected during a past
activity of the AGN, only ``old'' populations of bubbles which are
buoyantly rising are present.  It is likely that the cool blobs which
were tied to the radio bubbles when the AGN was active there, have
thermalized, the magnetic fields slow down but do not stop entirely
thermalization so that only a single phase gas is detected in the halo
regions.

\section{Summary}
\label{sec:concl}

The crisis of the standard cooling flow model brought about by {\it
  Chandra} and \XMM\, observations of galaxy clusters, has led to the
development of several heating models with the aim of identifying a
mechanism able to quench the cooling flow.

We have used observations of Virgo/M87 to inspect the dynamics of the
gas in the center of the cluster, and to study the heating processes
able to balance radiative losses.  We combined the observations of
three satellites, namely \XMM, {\it Chandra} and {\it Beppo-SAX}. By
means of the spectral deprojection technique, we inferred the profiles
for the temperature $T$ and the electron density $n_e$ for M87. The
temperature profile drops by a factor of $\sim 2$ in the inner
($\simlt 15$ kpc). The electron density $n_e$ is well described by a
2-$\beta$ profile.  Starting from these profiles, we derived some
related physical quantities such as the gravitational mass and the
entropy profile.  The gravitational mass profile shows a plateaux at
$\sim 10- 15 $ kpc, which could correspond to the edge of the central
cD.

In agreement with \citet{Voigt} and \citet{VF04} results on a set of
cluster, we found that the thermal conduction in M87 can balance the
cooling flow only in the outer part of the core (say, $r \simgt 15$
kpc), while in the inner $10-15$ kpc, where the temperature profile
becomes flatter and the conduction is no longer efficient, some
extra-heating is required.  We have determined the extra-heating
needed to balance the cooling flow in M87, assuming a thermal
conduction efficiency $f=0.3$ with respect to the Spitzer value, and
an outflow mass rate $\dot M = 1.6{\rm M_\odot /yr}$.  The high
quality of our combined dataset allows us to inspect properly the
innermost regions, deriving an accurate profile for the extra-heating
term $\cal H$ in the regions where thermal conduction is not
sufficient to quench the cooling flow.

Several models and simulations concerning heating mechanisms through
buoyant gas in the cluster ICM have been recently proposed in the
literature \citep[e.g.][]{RB02,beg01,ch02,Bohr02,BKa,BKb,Brug02}. We
have assumed that the heating is provided by the central AGN by means
of deposition of buoyant bubbles in the ICM according to the model
proposed by \citet{RB02}. The bubbles rise through the ICM and
expand doing work on the surrounding plasma and heating it up.  We
fitted the extra-heating required in M87, with the heating functions
proposed by RB02 (see eqs.  \ref{eq:hbr} - \ref{eq:qint}).  The RB02
model seems suitable to describe the behavior of the data.  By fixing
the adiabatic index $\gamma_b=4/3$ (relativistic bubbles), we find a
scale radius $r_0=4.39 \pm 1.83$ kpc and $A = (8.35 \pm 1.65) \times
10^{-24} \, {\rm erg \, s^{-1} }$, which corresponds to a
central AGN luminosity $ L = 5.95 \times 10^{42} \, {\rm erg \, s^{-1}
  }$.  The scale radius is of the order of the extension of the AGN
jet and the inferred AGN luminosity is similar to the one estimated by
\citet{owen} and \citet{DiM02} $L \sim 3-4 \times 10^{42} {\rm erg
  /s^{-1}}$. Smaller conductivity efficiencies ($f=0.1$) or larger
outflow mass rates ($\dot M = 16 {\rm M_{\odot}/yr}$) provide AGN
luminosities ($L \sim few 10^{43} {\rm \, erg \, s^{-1} }$) which are
somewhat higher than the estimations of \citet{owen} and
\citet{DiM02}.  However, if the efficiency of the thermal conduction
is reduced ($f=0.1$) the model functions of RB02 seem no longer
suitable to describe the heating needed to balance the cooling flow.
For higher conduction efficiencies ($ f = 1 $) or smaller outflow
rates values ($ \dot M = 0.16 {\rm M_{\odot}/yr}$) the inferred
luminosity is slightly reduced ($L \sim 3-4 \times 10^{42} { \rm \,
  erg /s^{-1}}$).  The different values derived for $L$ are of the
same order of the luminosity measures for the AGN luminosity in M87
obtained by \citet{owen} and \citet{DiM02}.  In all the cases
considered, the inferred scale radius $r_0$, which fixes the radius at
which bubbles are deposited and start rising, is $r_0 \sim 4-5$ kpc
which approximatively corresponds to the AGN jet extension.

Finally, we discussed a scenario where the bubbles are filled with
relativistic particles and magnetic field, responsible for the radio
emission in M87.  In this scenario the density contrast $\Delta$ is
expected to be as large as 1. The buoyant velocity $v_b$ of the
bubbles can be derived by balancing the buoyant force with the drag
force. For small bubbles, the rise velocity is subsonic.  Under the
hypothesis of equipartition between relativistic particles and
magnetic field in the bubbles and of equilibrium pressure with the
thermal ICM, we evaluated the filling factor $\phi$ of the radio
bubbles. We find $\phi \sim 0.01$ which is of the same order of the
filling factor of the cool thermal component observed in the regions
of the radio lobes (M02). Hence, we suggest that this cool thermal
component is structured in blobs tied to the radio bubbles. The
thermal conduction, which should rapidly thermalized the cool blobs,
is suppressed by the magnetic fields of the radio bubbles. The density
contrast of the buoyant bubble+blob system is $\Delta < 1$ further
reducing the rising velocity.
  
The radio galaxies are likely to be intermittent on a timescale $t_i
\sim 10^4-10^5$ yr and they are supposed to heat the ICM through a
succession of outbursts. During each outburst a population of radio
bubbles is injected into the ICM. The bubbles rise buoyantly in the
intracluster gas, heating it up.  The outbursts follow each other on
small timescale $t_i$ which is much shorter than the rise timescale
$t_{rise} \simgt 10^8$ yr of the bubbles so that the mechanism is
isotropic throughout the cluster and approaches the steady state.
  
The X--ray cool blobs are detected in the radio lobes where the
injection of radio bubbles is occurring at the present time. The blobs
are tied to the radio bubbles so that the thermal conduction is highly
suppressed by the magnetic fields. In the radio halo, the radio
bubbles injection occurred in the past. The radio bubbles are
buoyantly rising in the cluster atmosphere and it is likely that the
X--ray cool blobs which were tied to the bubbles when the AGN was
active in those directions, have in the meanwhile thermalized so that
only a single phase thermal component is present here.

\acknowledgments

The authors wish to thank Luigina Feretti and Mariachiara Rossetti for
useful discussions.  The authors are pleased to acknowledge the
referee C. Kaiser whose useful comments and suggestions have
significantly improved the paper.

\appendix
\section{Correction factor in the deprojection recipe for the background emission}
\label{appen}

By using eq. \eqn{eq:depro} the deprojected variable $f_{shell}$ is
obtained from $F_{ring}$ by subtracting off the contribution of the
outer shells, starting from the outermost annulus and moving inwards.
This basic prescription assumes that there is no emission beyond the
maximum radius $R_{max}$. Thus, the basic recipe \eqn{eq:depro} must
be corrected to account for a contribution to the X--ray spectra from
the gas beyond $R_{max}$.

In practice, in eq.  \eqn{eq:depro} the 2-D variable to be deprojected
$F_{ring}$ is replaced by an effective one $F^{eff}_{ring}$ defined
by:

\begin{equation}
F^{eff}_{ring}=F_{ring} - g_{ring} \cdot F_n \cdot A_{ring}/A_n,
\label{eq:feff}
\end{equation}

\noindent
where $A_{ring}$ is the area of the ring, $F_{ring}$ the variable to
be deprojected and $g_{ring}$ the correction factor of the ring.
$A_n$ and $F_n$ are the area and the variable to be deprojected of the
outer ring.

In order to determine $g_{ring}$, some assumptions on the shape at
large radii for the different {\it deprojected} quantities must be
made.  The classic solution consists in assuming that all $\varepsilon
T_{shell}$, $\varepsilon$ and $n_e$ have an $f \propto r^{-\alpha}$
dependence at large radii, with $\alpha=4$.  For this standard
assumption, the correction factor $g_j$ can be determined analytically
\citep{McL99}.

Hence, the correction to the $j$-th ring for the contribution
coming from the outer part of the cluster is:

\begin{equation}
g_j= {{R_n^2-R_{n-1}^2}\over {R_j^2-R_{j-1}^2}}
{{\int_{R_{j-1}}^{R_j} db\, b\, 
\int_{\sqrt{R_n^2-b^2}}^{+ \infty}{dz \over (b^2+z^2)^{\alpha/2}}} 
\over {\int_{R_{n-1}}^{R_n}db \, b\, 
\int_0^{+ \infty}{dz \over (b^2+z^2)^{\alpha/2}}}}.
\label{eq:gj}
\end{equation}

\noindent
$R$ and $b$ refer to the 2-D radii of the rings and $z$ is the
line-of-sight integration variable and $f \propto r^{-\alpha}$ has
been used. Of course $r^2=b^2+z^2$ holds.  

However, the standard assumption $\alpha=4$ is quite simplistic and
holds only for an isothermal cluster with $n_e \propto r^{-2}$ at
large radii (e.g. in the usual $\beta$-model with $\beta=2/3$).  On
the contrary, we prefer considering the dependence $r^{-\alpha}$ with
a generic $\alpha$, different for each deprojected quantity.  While
for $\alpha=4$ the correction factor could be calculated analytically,
for a generic $\alpha$ it must be determined numerically.  We
evaluated the $g_j$ factor of eq.  \eqn{eq:gj} by truncating the
integrals at a $10 R_{max}$ external radius, with steps $ 0.01R_{max}$
wide.  We verified that the numerical method for the simple case
$\alpha = 4$ provides results which differ by no more than a few
percent from the analytical values, and in any case the effect is
always limited to the outermost bins. In order to find out the correct
$\alpha$ for each quantity to be deprojected, we applied iteratively
the deprojection. We started from reasonable values of $\alpha$. Then
we applied the deprojection and derive the new $\alpha$ values from
the slopes of the deprojected profiles at large radii. We used these
new $\alpha$ values to determine the correction factor of eq.
\eqn{eq:gj} and applied again the deprojection working out new
$\alpha$ values.  We stopped when all the new $\alpha$ values differed
from the starting values by less than 4$\%$.

\clearpage
\newpage

\clearpage

\begin{table}
\caption{Emission-weighted temperatures in keV and
  Normalized Emission Integral (NEI) per unit area; $NEI$ is given
  in {\sc{xspec}} units, i.e.  $NEI = {{10^{-14}} \over {4\pi
      d^2_{ang}(1+z)^2}}EI$, where $d_{ang}$ is the angular distance
  of M87 in cm, $z$ the redshift and $EI$ in cm$^{-3}$. The $Area$ is
  in arcmin$^2$.\label{tab:t2d-ei}}
\begin{center}
\begin{tabular}{l|ccc}
\hline\hline
&$ r(arcmin) $ & $T_{EW}$(keV) & $NEI/Area$ \\
\hline
& 0.00- 0.17& 1.61$^{+0.06}_{-0.05}$& 13.84$^{+1.090}_{-1.024}\cdot 10^{-3}$\\
& 0.17- 0.50& 1.75$^{+0.06}_{-0.06}$& 10.83$^{+0.687}_{-0.746}\cdot 10^{-3}$\\
& 0.50- 0.75& 1.77$^{+0.05}_{-0.05}$&  6.84$^{+0.380}_{-0.362}\cdot 10^{-3}$\\
{\it Chandra} & 0.75- 1.00& 1.79$^{+0.03}_{-0.03}$&  4.99$^{+0.135}_{-0.138}\cdot 10^{-3}$\\
& 1.00- 1.33& 1.88$^{+0.04}_{-0.04}$&  3.50$^{+0.082}_{-0.078}\cdot 10^{-3}$\\
& 1.33- 1.67& 1.95$^{+0.05}_{-0.04}$&  2.60$^{+0.078}_{-0.076}\cdot 10^{-3}$\\
& 1.67- 2.00& 2.06$^{+0.03}_{-0.04}$&  2.40$^{+0.064}_{-0.064}\cdot 10^{-3}$\\
\hline
& 0.00- 0.50& 1.64$^{+0.01}_{-0.01}$& 14.63$^{+0.211}_{-0.199}\cdot 10^{-3}$\\
& 0.50- 1.00& 1.72$^{+0.02}_{-0.02}$&  4.96$^{+0.099}_{-0.100}\cdot 10^{-3}$\\
& 1.00- 1.50& 1.78$^{+0.02}_{-0.02}$&  3.26$^{+0.069}_{-0.070}\cdot 10^{-3}$\\
& 1.50- 2.00& 1.83$^{+0.03}_{-0.02}$&  2.58$^{+0.066}_{-0.072}\cdot 10^{-3}$\\
& 2.00- 2.50& 1.95$^{+0.02}_{-0.02}$&  2.21$^{+0.041}_{-0.041}\cdot 10^{-3}$\\
{\it XMM-Newton} & 2.50- 3.00& 2.01$^{+0.03}_{-0.03}$&  1.84$^{+0.040}_{-0.043}\cdot 10^{-3}$\\
& 3.00- 3.50& 2.00$^{+0.03}_{-0.03}$&  1.49$^{+0.043}_{-0.040}\cdot 10^{-3}$\\
& 3.50- 4.00& 2.06$^{+0.03}_{-0.03}$&  1.30$^{+0.028}_{-0.025}\cdot 10^{-3}$\\
& 4.00- 6.00& 2.17$^{+0.05}_{-0.04}$&  0.92$^{+0.015}_{-0.014}\cdot 10^{-3}$\\
& 6.00- 8.00& 2.26$^{+0.02}_{-0.02}$&  0.64$^{+0.007}_{-0.007}\cdot 10^{-3}$\\
& 8.00-10.00& 2.31$^{+0.02}_{-0.03}$&  0.45$^{+0.005}_{-0.005}\cdot 10^{-3}$\\
&10.00-13.00& 2.31$^{+0.02}_{-0.03}$&  0.30$^{+0.004}_{-0.004}\cdot 10^{-3}$\\
\hline
& 0.00- 2.00& 2.00$^{+0.04}_{-0.04}$&  5.05$^{+0.232}_{-0.217}\cdot 10^{-3}$\\
& 2.00- 4.00& 2.19$^{+0.04}_{-0.04}$&  1.71$^{+0.053}_{-0.056}\cdot 10^{-3}$\\
& 4.00- 6.00& 2.36$^{+0.05}_{-0.04}$&  0.94$^{+0.026}_{-0.027}\cdot 10^{-3}$\\
{\it Beppo-SAX} & 6.00- 8.00& 2.47$^{+0.07}_{-0.08}$&  1.01$^{+0.037}_{-0.037}\cdot 10^{-3}$\\
& 8.00-12.00& 2.63$^{+0.09}_{-0.09}$&  0.34$^{+0.015}_{-0.015}\cdot 10^{-3}$\\
&12.00-16.00& 2.60$^{+0.10}_{-0.10}$&  0.29$^{+0.014}_{-0.012}\cdot 10^{-3}$\\
&16.00-20.00& 2.41$^{+0.09}_{-0.09}$&  0.23$^{+0.012}_{-0.011}\cdot 10^{-3}$\\
\hline
\end{tabular}
\end{center}
\end{table}

\clearpage

\begin{table}
\caption{Temperature, electron density and emissivity values obtained with the 
deprojection technique for M87. \label{tab:tneps}} 
\begin{center}
\begin{tabular}{cccc}
\hline \hline
$r$ & $T$ & $n_e$ & $\varepsilon$ \\
(kpc)& (keV) & $10^{-3}$ cm$^{-3}$ & $10^{-28}$ erg s$^{-1}$ cm$^{-3}$ \\
\hline
  0.53&0.86$^{+0.64}_{-1.27}$&108.08$^{+27.33}_{-27.66}$&730.41$^{+727.80}_{-600.21}$\\
  2.10&1.66$^{+0.16}_{-0.14}$& 85.30$^{+ 6.48}_{- 7.58}$&762.65$^{+136.21}_{-138.51}$\\
  3.92&1.65$^{+0.15}_{-0.13}$& 54.50$^{+ 4.38}_{- 4.71}$&333.02$^{+ 64.25}_{- 56.44}$\\
  5.48&1.65$^{+0.08}_{-0.08}$& 45.30$^{+ 1.84}_{- 1.81}$&235.43$^{+ 26.01}_{- 23.21}$\\
  7.83&1.66$^{+0.08}_{-0.09}$& 25.04$^{+ 1.15}_{- 1.26}$& 68.63$^{+  7.47}_{-  7.20}$\\
 10.97&1.48$^{+0.12}_{-0.13}$& 18.77$^{+ 0.96}_{- 1.25}$& 34.24$^{+  4.47}_{-  4.39}$\\
 14.10&1.72$^{+0.12}_{-0.12}$& 16.77$^{+ 0.80}_{- 0.81}$& 25.93$^{+  2.67}_{-  2.48}$\\
 17.23&2.01$^{+0.22}_{-0.22}$& 15.38$^{+ 0.76}_{- 0.73}$& 12.62$^{+  2.82}_{-  3.26}$\\
 20.37&1.73$^{+0.18}_{-0.20}$& 11.47$^{+ 0.79}_{- 0.83}$& 12.47$^{+  2.00}_{-  1.93}$\\
 23.50&1.86$^{+0.15}_{-0.14}$& 12.24$^{+ 0.44}_{- 0.50}$& 14.99$^{+  1.29}_{-  1.21}$\\
 31.33&2.00$^{+0.15}_{-0.15}$&  7.44$^{+ 0.17}_{- 0.16}$&  4.64$^{+  0.75}_{-  0.69}$\\
 43.86&2.14$^{+0.10}_{-0.11}$&  6.53$^{+ 0.18}_{- 0.17}$&  4.45$^{+  0.46}_{-  0.41}$\\
 62.66&2.49$^{+0.36}_{-0.37}$&  2.81$^{+ 0.24}_{- 0.26}$&  0.90$^{+  0.18}_{-  0.20}$\\
 87.73&2.67$^{+0.31}_{-0.28}$&  2.50$^{+ 0.19}_{- 0.22}$&  0.75$^{+  0.12}_{-  0.12}$\\
112.79&2.37$^{+0.10}_{-0.10}$&  2.31$^{+ 0.06}_{- 0.06}$&  0.54$^{+  0.03}_{-  0.03}$\\
\hline
\end{tabular}
\end{center}
\end{table}


\begin{thebibliography}{}
  
\bibitem[Allen et al.(1996)]{Allen96} Allen, S. W., Fabian, A. C.,
  Edge, A. C., Bautz, M. W., Furuzawa, A., Tawara, Y., 1996, MNRAS,
  283, 263
  
\bibitem[Allen et al.(2001)]{Allen01} Allen, S.W. et al., 2001, MNRAS,
  324, 842

\bibitem[Begelman(2001)]{beg01} Begelman, M.C., 2001, in {\it Gas and
    Galaxy Evolution}, ASP Conf. Proc., vol. 240, ed. Hibbard, J.E.,
  Rupen, M.P., and van Gorkom, J.H., p. 363, (astro-ph/0207656)
  
\bibitem[Belsole et al.(2001)]{belsole} Belsole, E., Sauvageot, J.
  L., B\"ohringer, H., Worrall, D. M., Matsushita, K., Mushotzky, R.
  F., Sakelliou, I., Molendi, S., Ehle, M., Kennea, J., Stewart, G.,
  Vestrand, W. T., 2001, \aap, 365, L188
  
\bibitem[Binney and Cowie(1981)]{BC81} Binney, J. and Cowie, L.L.,
  1981, ApJ, 247, 464

\bibitem[B\"{o}hringer et al.(1994)]{bohr94} B\"{o}hringer, H., Briel, U. G.,
  Schwarz, R. A., Voges, W., Hartner, G., Trumper, J., 1994, Nature,
  368, 828

\bibitem[B\"{o}hringer et al.(2001)]{Bohr01} B\"{o}hringer, H.,
  Belsole, E., Kennea, J., Matsushita, K., Molendi, S., Worrall, D.,
  Mushotzky, R. F., Ehle, M., Guainazzi, M., Sakelliou, I., Stewart,
  G., Vestrand, W. T., Dos Santos, S., 2001, \aap, 365, L181
  
\bibitem[B\"{o}hringer et al.(2002)]{Bohr02} B\"{o}hringer, H.,
  Matsushita, K., Churazov, E., Ikebe, Y., Chen, Y., 2002, \aap, 382,
  804
  
\bibitem[Brighenti and Mathews(2002)]{BM02} Brighenti, F. and
  Mathews, W.G., 2002, ApJ, 573, 542
  
\bibitem[Br\"{u}ggen and Kaiser(2001)]{BK01} Br\"{u}ggen, M. and Kaiser,
  C.R., 2001, MNRAS 325, 676

\bibitem[Br\"{u}ggen and Kaiser(2002a)]{BKa} Br\"{u}ggen, M. and Kaiser,
  C.R., 2002, Nature, 418, 301
  
\bibitem[Br\"{u}ggen and Kaiser(2002b)]{BKb} Br\"{u}ggen, M. and Kaiser,
  C.R., 2002, MNRAS, 325, 676
  
\bibitem[Br\"{u}ggen et al.(2002)]{Brug02} Br\"{u}ggen, M., Kaiser,
  C.R., Churazov, E. and En{\ss}lin, T.A., 2002, MNRAS, 331, 545
  
\bibitem[Burns(1990)]{Burns} Burns, J.O., 1990, AJ, 99, 14

\bibitem[Burns, Owen and Rudnick(1979)]{bor} Burns, J.O.,Owen, F.N. and
  Rudnick,, L., 1979, AJ, 84, 1683
 
\bibitem[Chandran and Cowley(1998)]{CC98} Chandran, B.D.G. and
  Cowley, S.C., 1998, Phys. Rev. Lett., 80, 3077

\bibitem[Churazov et al.(2000)]{ch00} Churazov, E.,
  Forman, W., Jones, C. and B\"{o}hringer, H., 2000, \aap, 356, 788

\bibitem[Churazov et al.(2001)]{ch01} Churazov, E.,Br\"{u}ggen, M., Kaiser,
  C.R., B\"{o}hringer, H. and Forman, W., 2001, ApJ,  554, 261

\bibitem[Churazov et al.(2002)]{ch02} Churazov, E., Sunyaev, R.,
  Forman, W. and B\"{o}hringer, H., 2002, MNRAS, 332, 729
  
\bibitem[Clarke, Kronberg and B\"ohringer(2001)]{CKB} Clarke, T.
  E., Kronberg, P. P. and B\"ohringer, H., 2001, ApJ, 547, L111
     
\bibitem[De Grandi and Molendi(2001)]{DM01} De Grandi, S. and
  Molendi, S., 2001, ApJ, 551, 153

\bibitem[Di Matteo et al.(2003)]{DiM02} Di Matteo, T., Allen, S.W.,
  Fabian, A.C., Wilson, A.S., Young, A.J., 2003, ApJ, 582, 133

\bibitem[En{\ss}lin and Heinz(2002)]{EH02} En{\ss}lin, T.A. and 
  Heinz, S., 2002, \aap, 384, L27

\bibitem[Ettori(2002)]{ES02} Ettori, S., 2002, MNRAS, 330, 971
  
\bibitem[Ettori, De Grandi and Molendi(2002)]{EDM02} Ettori, S., De
  Grandi, S. and Molendi, S., 2002, \aap, 391, 841
  
\bibitem[Ettori et al.(2002)]{Ett02} Ettori, S., Fabian, A.C., Allen,
  S.W. and Johnstone, R.M., 2002, MNRAS, 331, 635
  
\bibitem[Fabian(1994)]{fa94} Fabian, A.C., 1994, ARAA, 32, 277
  
\bibitem[Fabian et al.(2001)]{Fab01} Fabian, A.C., Mushotzky, R. F.,
  Nulsen, P. E. J., Peterson, J. R., 2001, MNRAS, 321, L20
  
\bibitem[Fabian et al.(2002)]{Fab02} Fabian, A.C., Voigt, L.M. and
  Morris, R.G., 2002, MNRAS, 335, 71
  
\bibitem[Feretti(1999)]{F99} Feretti, L. 1999, in {\it Diffuse Thermal
    and Relativistic Plasma in Galaxy Clusters}, ed. H.
  B\"{o}hringer, L.  Feretti and P. Schuecker (MPE Rep. 271), 3
 
\bibitem[Feretti et al.(1999)]{Fer99} Feretti, L., Dallacasa, D.,
  Govoni, F., et al. 1999, \aap, 344, 472
  
\bibitem[Fusco-Femiano et al.(2000)]{fusco} Fusco-Femiano, R., Dal
  Fiume, D., Feretti, L., Giovannini, G., Grandi, P., Matt, G.,
  Molendi, S., Santangelo, A., 1999, ApJ, 513, L21

\bibitem[Gastaldello and Molendi(2002)]{GM02} Gastaldello, F. and
  Molendi, S., 2002, ApJ, 572, 160
  
\bibitem[Ghizzardi(2001)]{PSF1} Ghizzardi, S., 2001, 
 XMM-SOC-CAL-TN-0022 at \\
http://xmm.vilspa.esa.es/external/xmm\_sw\_cal/calib/documentation.shtml\#XRT

\bibitem[Giovannini and Feretti(2000)]{GF00} Giovannini, G. and
    Feretti, L., 2000, NewA, 5, 335

\bibitem[Gruzinov(2002)]{gruz02} Gruzinov, A., 2002, astro-ph/0203031
  
\bibitem[Gull and Northover(1973)]{GN73} Gull, S.F. and Noorthover, J.E., 
  1973, Nature, 244, 80

\bibitem[Hines, Owen and Eilek(1989)]{HOE} Hines, D.C., Owen, F.N. and
  Eilek, J.A., 1989, ApJ, 347, 713

\bibitem[Kaastra et al.(2001)]{Kaa01} Kaastra, J.S., et al., 2001,
  \aap, 365, L99

\bibitem[Kaiser(2003)]{Kai03} Kaiser, C.R., 2003, MNRAS, 343, 1319
  
\bibitem[Kim, Kronberg and Tribble(1991)]{kim91} Kim, K.T., Kronberg,
  P.P. and Tribble, P.C., 1991, ApJ, 379, 80

\bibitem[Kriss, Cioffi and Canizares(1983)]{Kriss83} Kriss, G. A.,
  Cioffi, D. F. and Canizares, C. R., 1983, ApJ, 272, 439

\bibitem[Malyshkin(2001)]{Mal01}   Malyshkin, L., 2001, ApJ, 554, 561

\bibitem[Markevitch(2002)]{Mark02} Markevitch, M., 2002,
  astro-ph/0205333
  
\bibitem[Matsushita et al.(2002)]{matsu} Matsushita, K., Belsole, E.,
  Finoguenov, A. and B\"ohringer, H., 2002, \aap, 386, 77
  
\bibitem[McLaughlin(1999)]{McL99} McLaughlin, D.E., 1999, AJ, 117,
  2398
  
\bibitem[McNamara et al.(2001)]{McN01} McNamara, B.R., et al., 2001,
  ApJ, 562, L149
  
\bibitem[McNamara(2002)]{McN02} McNamara, B.R., 2002, ``{\it The
    High-Energy Universe at Sharp Focus: Chandra Science}'',
  astro-ph/0202199
 
\bibitem[Molendi(2002)]{ms02} Molendi, S., 2002, ApJ, 580, 815  (M02)
  
\bibitem[Molendi and Gastaldello(2001)]{MG01} Molendi, S. and
  Gastaldello, F., 2001, \aap, 375, L14
  
\bibitem[Molendi and Pizzolato(2001)]{MP01} Molendi, S. and
  Pizzolato, F., 2001, ApJ, 560, 194
  
\bibitem[Narayan and Medvedev(2001)]{NM01} Narayan, R. and Medvedev,
  M.V., 2001, ApJ, 562, L129

\bibitem[Nulsen and B\"ohringer(1995)]{NB95} Nulsen, P.E.J. and
  B\"ohringer, H., 1995, MNRAS, 274, 1093
  
\bibitem[O'Dea and Owen(1987)]{OO} O'Dea, C.P. and Owen, F.N., 1987,
  ApJ, 316, 95
  
\bibitem[Owen, Eilek and Kassim(2000)]{owen} Owen, F.N., Eilek, J.A. and Kassim,
  N.E., 2000, ApJ, 543, 611 (OEK)
  
\bibitem[Owen, Morrison and Voges(1999)]{OMV} Owen, F. N., Morrison,
  G., and Voges, W., 1999, in {\it Diffuse Thermal and Relativistic Plasma
  in Galaxy Clusters}, ed. H. B\"{o}hringer, L. Feretti and  P. Schuecker
  (MPE Report 271), 9
  
\bibitem[Pacholczyk(1970)]{Pacho70} Pacholczyk,A.G., 1970, {\it Radio
    Astrophysics} (San Francisco: Freeman)
  
\bibitem[Peterson et al.(2001)]{Pet01} Peterson, J. R., et al., 2001,
  \aap, 365, L104
  
\bibitem[Pizzolato et al.(2003)]{pizzo02} Pizzolato, F., Molendi, S.,
  Ghizzardi, S., De Grandi, S., 2003, ApJ, 592, 62
  
\bibitem[Reynolds and Begelman(1997)]{RB97} Reynolds, C.S. and
  Begelman, M.C., 1997, ApJ, 487, L135
  
\bibitem[Ruszkowski and Begelman(2002)]{RB02} Ruszkowski, M. and
  Begelman, M.C., 2002, ApJ, 581, 223 (RB02)

\bibitem[Sarazin(1988)]{Sarazin} Sarazin, C.L., 1988, {\it X--ray
    emission from clusters of galaxies}, Cambridge University Press
  
\bibitem[Spitzer(1962)]{spitzer} Spitzer, L., 1962, {\it Physics of
    fully ionized gases}, New-York: Wiley Interscience
  
\bibitem[Tamura et al.(2001)]{Tam01} Tamura, T. et al., 2001, \aap,
  365, L87
  
\bibitem[Taylor et al.(2001)]{Tay01} Taylor, G. B., Govoni, F., Allen,
  S., and  Fabian, A. C., 2001, MNRAS, 326, 2

\bibitem[Voigt et al.(2002)]{Voigt} Voigt, L. M., Schmidt, R. W.,
  Fabian, A. C., Allen, S. W., Johnstone, R. M., 2002, MNRAS, 335, 7
  
\bibitem[Voigt and Fabian(2004)]{VF04} Voigt, L. M. and Fabian, A.C.,
  2004, MNRAS, 347, 1130
  
\bibitem[Young et al.(2002)]{Young} Young, A. J., Wilson, A. S. and
  Mundell, C. G., 2002, ApJ, 579, 560

\bibitem[Zakamska and Narayan(2003)]{ZN02} Zakamska, N.L. and Narayan R., 
  2003, ApJ, 582, 162

\end{thebibliography}
\end{document}